\documentclass[aps,twocolumn,superscriptaddress,pra,10pt,floatfix]{revtex4-2}
\usepackage{graphicx} 
\usepackage{newtxmath}
\usepackage{textcomp}
\usepackage{placeins}
\usepackage{ragged2e}

\begin{document}

\title{The Effect of Trap Design on the Scalability of Trapped-Ion Quantum Technologies}
\author{Le Minh Anh Nguyen}
\author{Brant Bowers}
\author{Sara Mouradian}
\affiliation{Electrical and Computer Engineering Department, University of Washington, Seattle, WA 98105}
\date{\today}

\begin{abstract}
To increase the power of a trapped ion quantum information processor, the qubit number, gate speed, and gate fidelity must all increase. All three of these parameters are influenced by the trapping field which in turn depends on the electrode geometry. Here we consider how the electrode geometry affects the radial trapping parameters: trap height, harmonicity, depth, and trap frequency. We introduce a simple multi-wafer geometry comprising a ground plane above a surface trap and compare the performance of this trap to a surface trap and a multi-wafer trap that is a miniaturized version of a linear Paul trap. We compare the voltage and frequency requirements needed to reach a desired radial trap frequency and find that the two multi-wafer trap designs provide significant improvements in expected power dissipation over the surface trap design in large part due to increased harmonicity. Finally, we consider the fabrication requirements and the path towards integration of the necessary optical control. This work provides a basis to optimize future trap designs with scalability in mind.
\end{abstract}

\maketitle

\section{Introduction}

\begin{figure*}
\includegraphics[width=\textwidth]{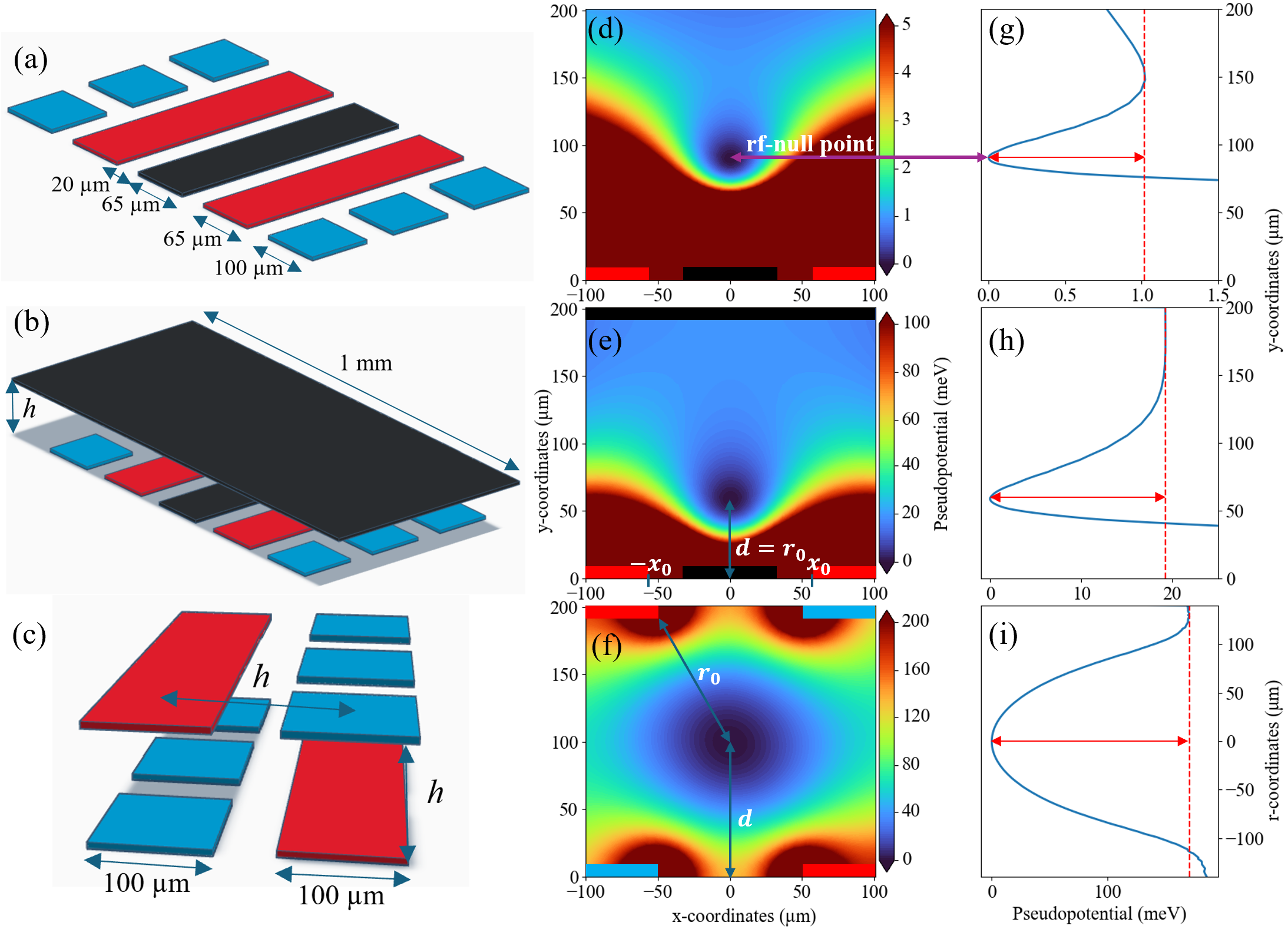}
    \caption{(a-c) electrode geometries for the surface trap, ``gnd-surface'' trap, and ``cross-rf'' trap, respectively. rf-electrodes are in red, dc-electrodes are in blue, and all grounded electrodes are in black. $h$ is the separation between the two wafers in the two multi-wafer designs and is varied to understand its impact on trap parameters. (d-f) 2D pseudopotential maps of the \textit{x-y} radial cross-section at $z=0$ of (a-c), respectively. (g-i) are the 1D pseudopotential slices at $z = 0$, $x=0$ for the structures in (a,b) and along $r=\sqrt{x^2+y^2}$ for the structure in (c). The red dashed line marks the pseudopotential saddle point used to calculate the trap depth.}
\label{Fig:TrapConfig}
\end{figure*}

The electronic and motional states of ions trapped in an electromagnetic field can be used to store and process quantum information, enabling a host of quantum technologies, such as atomic clocks~\cite{Ludlow_2015}, quantum sensors~\cite{Gilmore2021Aug}, and analog~\cite{Blatt2012Apr} and digital~\cite{Wineland_2003, Schindler_2013} quantum computers. Qubits encoded in the electronic states of trapped ions enjoy fast and high-fidelity state preparation and measurement (SPAM)~\cite{An_2022}, along with single- and two-qubit operations using optical~\cite{Ballance_2016, Clark_2021, Sawyer_2021} or electronic~\cite{Ospelkaus_2011, Weber_2024, Loschnauer_2024} gates. Traditionally, trapped-ion experiments use a macroscopic linear Paul trap with a deep and harmonic trapping potential~\cite{Paul_1990}. However, the bulky structure limits the number of qubits in a given volume. 

2D surface-electrode traps have emerged as an alternative to the traditional 3D Paul trap~\cite{Chiaverini_2005} and are currently the prevailing choice for trapped-ion experiments due to their versatility and scalability. Indeed, surface traps underpin the most advanced trapped-ion quantum information processors to date, enabling high-fidelity quantum computing demonstrations with tens of qubits~\cite{Zhang_2017, Pino_2021, Pogorelov_2021, Moses_2023, Chen_2024, Postler_2024} including initial demonstrations towards fault-tolerant encoding~\cite{Postler_2024} and operation~\cite{Reichardt_2024}. Still, the field remains far from a utility-scale trapped-ion quantum processor. Though many factors hinder the scalability of trapped-ion quantum technologies, here we focus on the effect of the trap electrode geometry. 

The geometry of the radio-frequency (rf) electrodes fixes the shape and relative strength of the trapping potential. Here we are interested in understanding trap geometry affects the trap frequency for a given set of drive parameters. While there are entangling gate schemes that do not use the motional modes~\cite{Zhang_2020,Saha2025Mar}, or that can be run arbitrarily fast~\cite{Saner_2023, Schafer_2018}, many platforms currently rely on gates with a speed limit set by the radial trap frequency~\cite{Steane_2000}. Moreover, a higher trap frequency will reduce gate errors due to electric field noise~\cite{Brownnutt_2015}. The requirement to trap many ions in potentials with high radial trap frequencies suggests that power dissipation may become a problem if not considered during trap optimization. Finally, any scalable trap geometry must be reproducible in an industrial setting and provide enough optical access for the necessary control beams. 

With these constraints in mind, we consider three distinct trap geometries for their suitability for scaling trapped-ion quantum systems. First, we consider a surface trap (Fig.~\ref{Fig:TrapConfig}(a)) with a representative electrode geometry~\cite{Chiaverini_2005}. Second, we consider a miniature version of a linear Paul trap (Fig.~\ref{Fig:TrapConfig}(c)), or ``cross-rf'' trap. Finally, we introduce a simpler multi-wafer trap design (Fig.~\ref{Fig:TrapConfig}(b)) which consists of the same surface trap but with an added ground plane above the electrodes. We do not fully optimize the trap geometries, but rather explore the effect of wafer separation on the trapping field for the two multi-wafer designs. We restrict our discussion to the scaling of the radial trapping potential, as this is completely defined during trap fabrication. In contrast, the axial trapping field is set by the voltages applied to the dc electrodes. Thus in the following we focus on understanding the scaling of trapping parameters to guide future work in trap optimization for particular experimental design constraints.


We calculate the expected pseudopotential, find the trapping height, and evaluate the three designs through three figures of merit: harmonicity $k$ (Section~\ref{Sec:Harmonicity}), trap depth $D$ (Section~\ref{Depth}), and radial trapping frequency $\omega_{\text{rad}}$ (Section~\ref{Sec:Freq}). We use these figures of merit to compare the expected performance and power dissipation of the three designs within a set of reasonable experimental assumptions (Section~\ref{example}). Finally, we discuss optical access and fabrication techniques (Section~\ref{Sec:Fab}).

\section{Field Characterization}
\label{Sec:Characterization}

\subsection{Calculation of Trapping Potential}
\label{Sec:Simulation}

The radial quadrupole electric potential due to an rf drive with amplitude $V_\text{rf}$, frequency $\Omega_\text{rf}$, and phase $\phi$ is~\cite{Ghosh_1995}:
\begin{equation}
    \Phi_\text{rf}(x,y,t) = \frac{V_\text{rf} }{2r_0^2}(k_\text{x} x^2 + k_\text{y} y^2)\cos(\Omega _\text{rf} t + \phi) 
    \label{Eqn:QuadrupoleField}
\end{equation}
where $r_0$ is the distance from the ion to the closest electrode surface and $k_\text{x}$ and $k_\text{y}$ are coefficients that satisfy the Laplace equation $\nabla^2 \Phi _\text{rf} = 0$.

Rapid oscillations of the rf field create a ponderomotive potential, or ``pseudopotential'',  proportional to the square of this field's amplitude, $|E(\textbf{r})|^2$.  For a particle with charge $e$ and mass $m$ in an oscillating electric field, the pseudopotential is~\cite{Kajita_2022}: 
\begin{equation}
     \psi(\textbf{r}) = \frac{e}{4m\Omega_\text{rf}^2}|\nabla \Phi_\text{rf}(\textbf{r})|^2 = \frac{e}{4m\Omega_\text{rf}^2}|E(\textbf{r})|^2 
     \label{Eqn:Pseudopotential}
\end{equation} 
In the following we consider the pseudopotential for $^{40}$Ca$^+$ although our results are directly applicable to any ion species. We assume a source with voltage $V_{\textrm{rf}}=10$\,V and frequency $\Omega_{\textrm{rf}}=2\pi\times20$\,MHz unless otherwise specified.

We simulate the electric field due to each rf electrode to obtain the pseudopotential for a given geometry. The simulations are done in COMSOL Multiphysics using the boundary element (BE) method. In the BE method, the boundary constraints defined by the electrodes are used to define the integral form of Maxwell's equations for the structure of interest. These equatiosn are tehn solved for points on a pre-defined mesh. This method has been used in a number of previous studies and has been validated against experiments~\cite{Pearson_2006, Allcock2010May, Hong_2016,Doret2012Jul,Moehring2011Jul,Wright2013Mar}. The full simulation volume is $8\times8\times5$\,mm$^3$. This area is meshed using a built-in adaptive meshing routine within COMSOL with a element size of 100\,\textmu m. A finer meshing volume of $0.6\times0.6\times0.5$\,mm$^3$ is centered at the expected rf-null point. The same COMSOL adaptive meshing algorithm is used with an element size of 1-10\,\textmu m giving a minimum mesh size of 1\,\textmu m at the ion location. Each simulation takes approximately half an hour to run. 


We construct the pseudopotential from teh simulated electric field following Eq.~\ref{Eqn:Pseudopotential}. Figs.~\ref{Fig:TrapConfig}(d-f) show the \textit{xy} cross-section of the pseudopotential of the three trap designs, reported in meV. For the two multi-wafer trap designs, we pick the same wafer separation, $h= 200$\,\textmu m to illustrate the qualitative differences in the pseudopotential for the different trap designs.  The ion is trapped where the pseudopotential is zero, as indicated. Figs.~\ref{Fig:TrapConfig}(g,h) show the pseudopotential along $y$ coordinate at $x=0$ for surface and gnd-surface geometries, while Fig.~\ref{Fig:TrapConfig}(i) is the pseudopotential running from rf electrode to rf electrode crossing through the rf-null point. The trap depth is indicated by the red arrow, and the dashed line illustrates the saddle point (see Section~\ref{Depth}).

In the following sections, the trapping performance of each trap geometry is investigated for varying wafer separation, $h$. For the cross-rf trap, the horizontal electrode-electrode spacing is also kept at $h$ to maintain the symmetry of the potential (see Fig.~\ref{Fig:TrapConfig}(c)). All other parameters are kept constant. 

\subsection{Ion Height}
The ion height, $d$, is the vertical distance between the rf-null point and the bottom wafer, as indicated in Figs.~\ref{Fig:TrapConfig}(d-f). The position of the rf null does not directly impact the scalability of a trapped-ion system, but it does influence the achievable trap frequency (Section~\ref{Sec:Freq}) and heating rate (Section~\ref{example}) and has implications for optical addressing~\cite{Mehta2016Dec,Maunz_2016}. We note that $d$ can also be controlled by changing the lateral geometry of the electrodes~\cite{Gerasin_2024}, but here we focus on the relationship between trapping height $d$ and wafer separation $h$ for the multi-wafer traps. 

\begin{figure}
    \centering
\includegraphics[width=\linewidth]{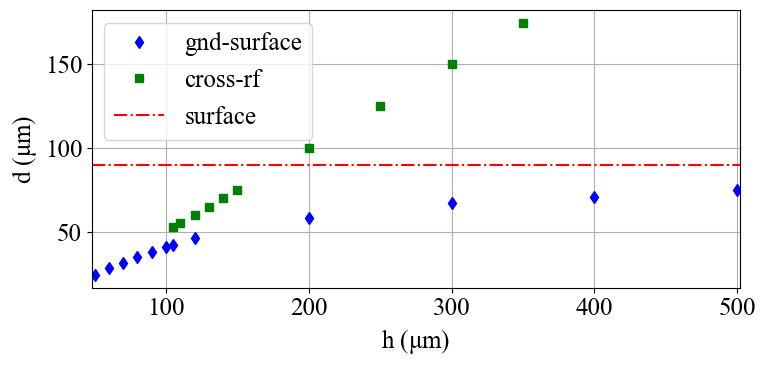}
    \caption{ Ion height $d$ as a function of wafer separation $h$ for the gnd-surface, and cross-rf traps. The ion height for the surface trap is shown for reference.}
    \label{fig:IonPos}
\end{figure}

Fig.~\ref{fig:IonPos} presents the ion height $d$ derived from simulations as a function of $h$, the distance between wafers. The trap height for the surface trap is shown in red for reference. Our uncertainty for the estimation of $d$ is defined by our mesh size of 1\,\textmu m. This uncertainty could be reduced with interpolation or a finer mesh. For the cross-rf trap, the rf-null point is always directly between the two wafers, or $d=h/2$. 
As expected, as $h$ increases the rf-null point of the gnd-surface trap approaches that of the bare surface trap.
\label{Sec:Position}

\subsection{Harmonicity}
\label{Sec:Harmonicity}
Harmonicity, $k$, measures how closely the trapping field resembles an ideal harmonic potential. From Eq.~\ref{Eqn:QuadrupoleField}, $k_\text{x} = -k_\text{y} = 1$ for a perfect quadrupole field. Deviations from this can cause gate errors due to increased cross-Kerr nonlinearity~\cite{Marquet_2003, Home2011Jul}, and the entangling gate control must be optimized with the particular anharmonicity in mind~\cite{Wu2018Jun}. Anharmonicity also reduces the achievable trap frequency for a given drive voltage in two ways. First, it reduces the amount of applied voltage  converted to a quadratic trapping potential as described below (Eq.~\ref{Eqn:frequency}). It also reduces the experimentally accessible extent of the stability region of the dynamic trapping field~\cite{Alheit_1997, Razvi_1998, Drakoudis_2006}. A full multipole expansion of the field can provide further insight into the expected ion dynamics and provide corrections to the trap frequencies~\cite{Littich}, but here we only consider harmonicity. 

\begin{figure}
    \includegraphics[width=\linewidth]{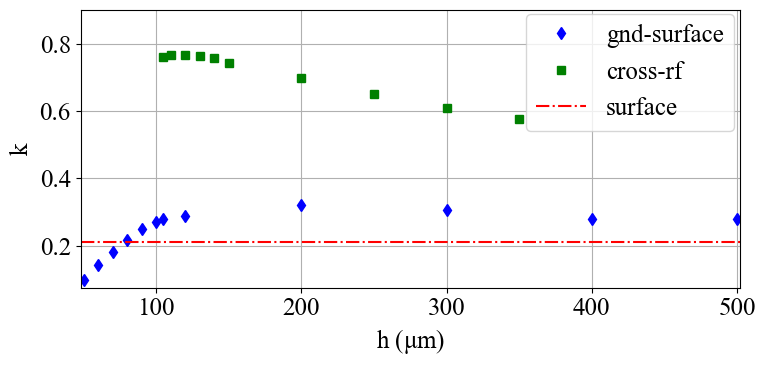}
    \caption{ Harmonicity $k$ as a function of wafer separation $h$ for the multi-wafer traps. The harmonicity for the surface trap design is shown for reference. }
    \label{fig:Harmonicity}
\end{figure} 

We determine radial harmonicity by performing a quadratic fit of the electric potential along the two radial axes. For the multi-wafer traps, we investigate how harmonicity changes with wafer separation $h$. The cross-rf trap is symmetric and we conducted the fit along the diagonal axis as describe in Fig.~\ref{Fig:TrapConfig}(i). For the surface and gnd-surface traps, $k_y < k_x$ due to the asymmetry of the $y$ axis. Fig.~\ref{fig:Harmonicity} shows $k_y$ as a function of $h$ for the two multi-wafer traps. The error bars derived from the quadratic fit are smaller than the symbol size and are all less than 0.001. 

The surface trap design (red line) suffers from large anharmonicity due to the highly non-symmetric geometry. The gnd-surface trap exhibits a modest increase in harmonicity in comparison to the surface trap for most heights. However, it suffers a steep decrease in $k$ when $h$ is less than the rf-rf electrode distance. For large separation, the configuration effectively transitions into a surface-electrode trap. 
In contrast, the harmonicity of the cross-rf trap is inversely proportional to $h$ 
and is generally high due to its symmetric geometry.



\subsection{Trap Depth}
\label{Depth}
The trap depth $D$ is the minimum potential energy difference between the rf-null point and the saddle point, which is indicated by a red arrow in Figs.~\ref{Fig:TrapConfig}(g-i). $D$ defines the maximum kinetic energy an ion can have without escaping confinement. Low trap depth can hinder ion loading and increase ion loss due to background gas collisions~\cite{Sage2012Jul,Hong_2016}. As above, we only consider the contribution from the rf trapping fields and do not consider the effect of the dc potential as that is determined post-fabrication by the voltages applied to the dc electrodes.

\begin{figure}[t]
    \centering
    \includegraphics[width=\linewidth]{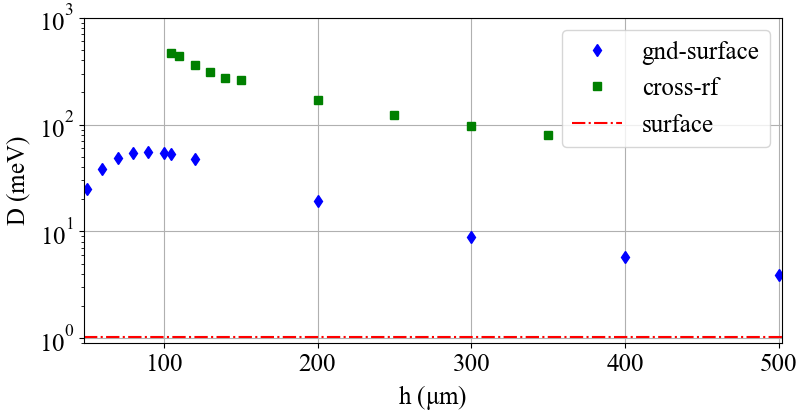}
    \caption{Trap depth $D$ as a function of wafer separation $h$ for the multi-wafer trap designs. $D$ for the surface trap design is shown for reference.}
    \label{fig:TrapDepth}
\end{figure}

Fig.~\ref{fig:TrapDepth}(a) shows the relationship between $D$ and $h$ for the multi-wafer traps. The trap depth for the surface trap design is shown in red. We use the same trapping parameters as in previous sections, $\Omega_{\textrm{rf}} = 2\pi\times20$\,MHz and $V_{\textrm{rf}}=10$\,V. 
For both multi-wafer traps $D$ generally increases with decreasing $h$. Though again the field of the gnd-surface trap is distorted when the wafer height is less than the rf-rf electrode separation, reducing $D$. 

\subsection{Trap frequency}
\label{Sec:Freq}

Finally, we discuss the radial trap frequency. The radial oscillation frequency of a charged particle with charge $e$ and mass $m$ with a drive voltage $V_\text{rf}$ and drive frequency $\Omega_\text{rf}$ is~\cite{Ghosh_1995}:
\begin{equation}
    \omega_\text{rad} =  \frac{V _\text{rf} k e}{\sqrt{2} m \Omega _\text{rf} r_0^2} = \frac{q\Omega _\text{rf}}{2\sqrt{2}} 
    \label{Eqn:frequency}
\end{equation}
where $q$ on the left-hand side of Eqn.~\ref{Eqn:frequency} is the instability parameter.
\begin{equation}
    q = \frac{2eV _\text{rf} k}{m\Omega _\text{rf} ^2r_0^2} = 2\sqrt{2}\frac{\omega_\text{rad}}{\Omega_\text{rf}}
    \label{Eqn:q}
\end{equation}

\begin{figure*}[hbtp]
    \includegraphics[width=0.7\textwidth]{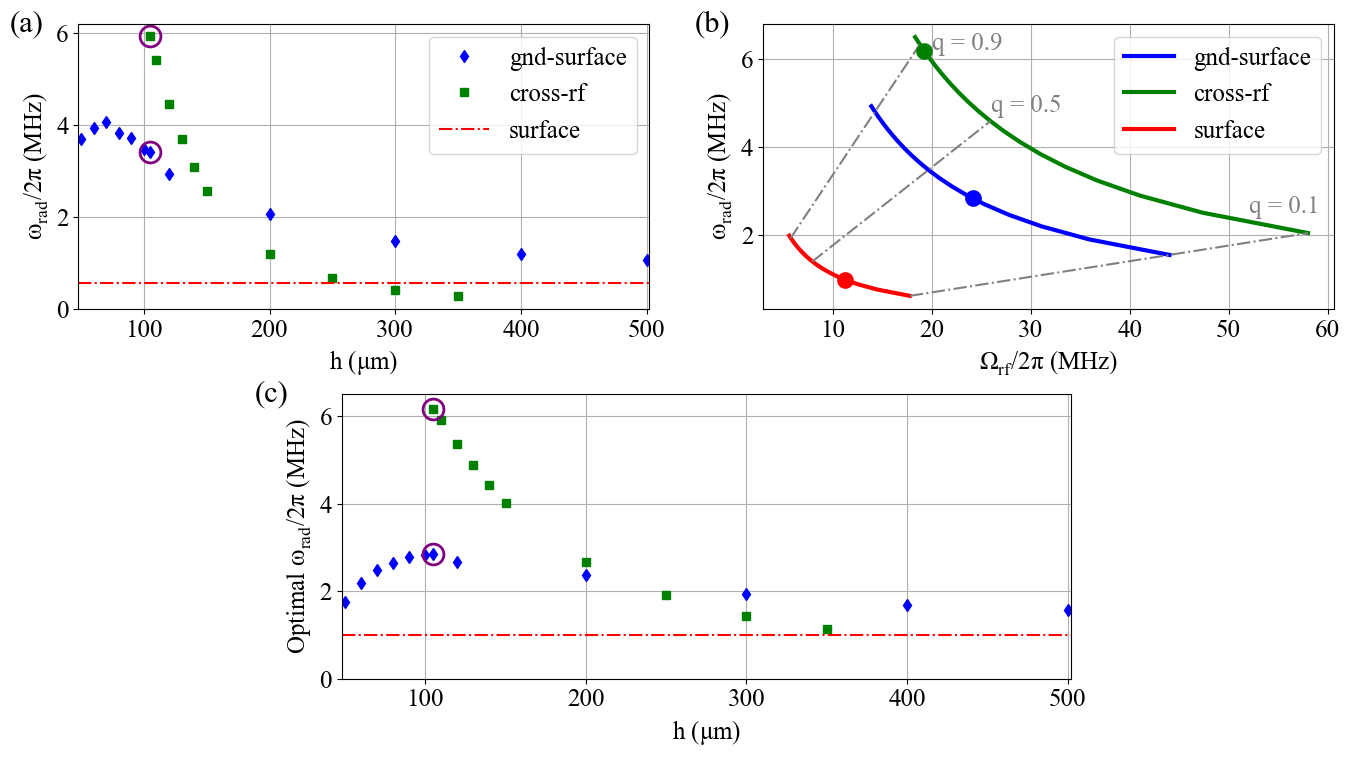}
    \caption{ (a)~Radial trap frequency as a function of wafer separation $h$ for the multi-wafer trap designs. The radial trap frequency for the surface trap design is shown for reference. $V_{\textrm{rf}} = 10$\,V and $\Omega_{\textrm{rf}} = 2\pi\times20$\,MHz. 
    (b)~Radial trap frequency as a function of $\Omega_\text{rf}$ at $V_{\textrm{rf}} = 10$\,V for the three trap geometries. The gray dashed lines indicate constant instability parameter $q$. The points denote the $q$ values used for each geometry in (c), as calculated in the text. (c)~Maximum radial frequency for a 10\,V drive assuming harmonicity-defined $q$-parameter as indicated in (b). The purple circles in (a) and (c) highlight the points at $h_\text{gnd-wafer} = h_\text{cross-rf} = 105$ \textmu m, discussed in Section~\ref{example}. }
    \label{fig:TrapFrequency}
\end{figure*}

In an ideal harmonic system, the first stability region of the Matthieu equation spans the entire range $0 \leq q \leq 1$~\cite{Ghosh_1995, Razvi_1998} and ideally $\Omega_{\text{rf}}$ and $V_{\text{rf}}$ would be chosen such that $q\simeq1$ increases the trap frequency achievable for a given voltage, as illustrated in Fig.~
\ref{fig:TrapFrequency}(b). However, for surface traps, the trap frequency is generally kept at $\omega < 0.1\Omega_\text{rf}$, or $q < 0.3$~\cite{Dehmelt_1968, Paul_1990} to achieve stable trapping. This deviation from the ideal case is attributed to the anharmonicity of the trap which is not captured by the idealized potential derived from the Matthieu Equation~\cite{Leibfried_2003}. Thus, the achievable radial trap frequency depends on the trap geometry through both the harmonicity $k$ and the trapping position $r_0$.

The cross-rf wafer trap is symmetric along the principle radial axes indicated in Fig.~\ref{Fig:TrapConfig}(i). For the surface and the gnd-surface wafer trap we simply focus on the radial frequency $\omega_y$ to be consistent with the axis we have chosen for harmonicity and trap depth.
Fig.~\ref{fig:TrapFrequency}(a) shows the radial frequency of a $^{40}$Ca$^+$ ion in the three trap designs for $\Omega_{\textrm{rf}}= 2\pi\times20$\,MHz  and $V_{\textrm{rf}}=10$\,V. As above, we vary the wafer separation $h$ for the two multi-wafer trap designs. The trap frequency for the surface trap under the same drive parameters is shown in red as a reference.

The trap frequency generally increases with smaller $h$, although trap frequency reduces when $h$ is less than the electrode-electrode separation for the gnd-surface trap. 

Fig.~\ref{fig:TrapFrequency}(b) shows the relationship between $\Omega_{\textrm{rf}}$, the $q$-parameter, and the trapping frequency for the three trap designs for $V_\text{rf} = 10$\,V.  For the two multi-wafer traps $h = 105$\,\textmu m as marked in Fig.~\ref{fig:TrapFrequency}(a). At the same $q$, the multi-wafer traps produce a higher radial trap frequency, with the cross-rf trap having a slightly better performance. High trap depth~\cite{Razvi_1998} and high harmonicity~\cite{Xu_2023} allow for certain traps to operate at much higher instability $q$ value, increasing the attainable radial trap frequency~\cite{Xu_2023}.

Trap frequency can be increased by increasing the trap drive amplitude,$V_\textrm{rf}$. However, the maximum voltage is set by the breakdown field of the dielectric used to separate neighboring electrodes~\cite{Sterling_2013}. Moreover, for large-scale trapped-ion systems, increased rf amplitude and frequency will increase power dissipation, which can be limiting, especially for systems working at cryogenic temperatures. This is discussed further in Section~\ref{example}. Therefore, it is important to understand the maximum trap frequency that can be achieved given a set drive amplitude.

Fig.~\ref{fig:TrapFrequency}(c) shows the maximum achievable frequency for $V_\textrm{rf} = 10$\,V) as a function of $h$. To find this, we pick an operating $q$ parameter for each trap by scaling a typical working parameter for surface traps ($q = 0.250$) with the harmonicity of each design, e.g. $q = q_{\textrm{st}} \frac{k}{k_{\textrm{st}}}$~\cite{Ghosh_1995}. This results in $q=0.905$ for the cross-rf trap, matching recent results in a miniature 3D-printed trap~\cite{Xu_2023}, and a more modest $q=0.333$ for the gnd-surface trap. These $q$ parameters are indicated in Fig.~\ref{fig:TrapFrequency}(b). This $q$ parameter then defines the drive frequency for a given $V_{\textrm{rf}}$ with Eq.~\ref{Eqn:q}. These parameters then give the maximum achievable trap frequency for a given rf drive amplitude from Eq.~\ref{Eqn:frequency}. As seen in Fig.~\ref{fig:TrapFrequency}(c), the multi-wafer traps show a gain in maximum trap frequency due to the increased harmonicity. In particular, the cross-rf trap provides a more than 10x improvement over the surface trap design considered here (6.12\,MHz vs 1.00\,MHz). For the gnd-surface trap in particular, we find that the maximum trap frequency (2.84\,MHz) is achieved when the wafer height equals the rf-rf electrode separation. These results can be used to focus design optimization for other electrode geometries and highlight the importance of the harmonicity of the trapping potential. 

\section{Expected Experimental Performance}
\label{example}

In this section we explore the expected experimental performance of the three trap designs. We use the field-dependent parameters presented above to understand the expected heating rate and how the necessary voltage and expected power dissipation scales with target trap frequency. 

\subsection{Heating Rate}
Noisy electric fields at frequencies near the secular trap frequency $\omega$ cause a time-dependent increase in the thermal population of an ion's motional modes~\cite{Ray_2019, Hite_2021} causing entangling gate errors~\cite{Cirac_2000}. The origin of electric field noise is not well understood~\cite{Brownnutt_2015} but the average increase in phonon number, or the heating rate $\dot{\overline{n}}$, generally fits the following trend~\cite{Brownnutt_2015}:
\begin{equation}
    \label{Eqn:heating}
    \dot{\overline{n}} \propto \omega^{-2}r_0^{-4}
\end{equation}

\begin{figure}
    \includegraphics[width=\linewidth]{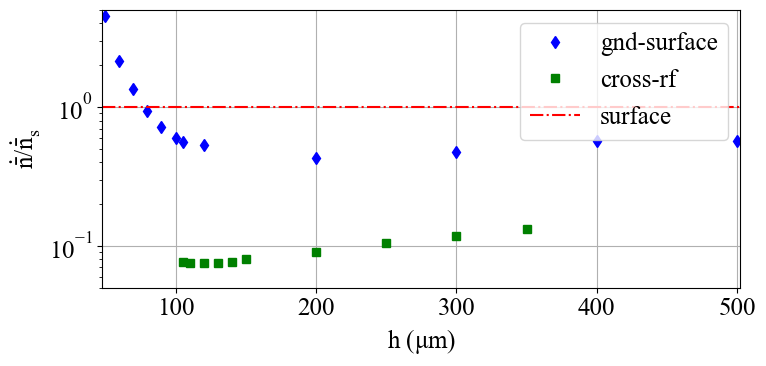}
    \caption{ The expected heating rate for the multi-wafer trap designs normalized to the expected heating rate of the surface trap. The normalized heating rate for the surface trap is shown for reference (red line).} \label{fig:Heating rate}
\end{figure}

Unfortunately, $\omega$ and $r_0$ generally counteract each other (Eq.~\ref{Eqn:frequency}). The expected heating rates of the three trap designs, normalized to the expected heating rate of a surface trap (Eq.~\ref{Eqn:heating}), are shown in Fig.~\ref{fig:Heating rate}. 
This indicates that with all other factors kept equal the cross-rf trap would enjoy nearly an order of magnitude lower heating rate than the other two designs and the gnd-surface trap can still achieve approximately half the heating rate of a surface trap. This may be negated by the introduction of more surface area, depending on the origin of the noise. Still, it is clear that multi-wafer traps can help reduce the heating rate under the assumption that the electric field noise is not affected by trap geometry.

\subsection{Achieving a Target Trap Frequency}

Here we compare the performance of the surface trap design and the two multi-wafer traps when $h=105$\,\textmu m (highlighted in Fig.~\ref{fig:TrapFrequency}). This wafer height provides both high harmonicity and large trap frequency for both multi-wafer trap designs. The rf-null position $d$ and harmonicity $k$ are set by the trap geometry. We use the operating $q$ parameters found in Section~\ref{Sec:Freq} and reported in Table~\ref{Table:example}. Here, we use the values of $d$, $k$, and $q$ to calculate the minimum trap voltage and drive frequency needed to achieve a target trap frequency of $\omega_{\textrm{rad}} = 2\pi\times10$\,MHz using Eq.~\ref{Eqn:frequency} and Eq~\ref{Eqn:q}. This is summarized in Table~\ref{Table:example}. This example underlines again the importance of harmonicity -- as in Fig.~\ref{fig:TrapFrequency}(c), a higher harmonicity reduces the necessary source voltage and frequency by increasing the workable $q$ parameter and by improving the conversion efficiency between applied voltage and trap frequency. Thus, the multi-wafer traps require significantly reduced drive voltage and frequency with respect to the surface trap geometry. This will be particularly important if trapped-ion systems are to increase the number of ions per chip.

The drive parameters define the expected power dissipation for each of the trap geometries. Power dissipation can occur due to electrical and dielectric losses. Here we focus on electrical power consumption as dielectric loss varies with material. COMSOL simulations show that the capacitance of the rf electrodes in the three trap designs are all around 0.1\,fF ($C_\text{surface} = 91.0$\,fF, $C_\text{gnd-surface} = 115.0$\,fF, and $C_\text{cross-rf} = 89.5$\,fF). In practice, the capacitance will likely be dominated by the trace layout and external wiring as standard values are in the 1-10\,pF range. Thus we present the power dissipation normalized to the system capacitance, which scales with the amplitude and frequency of the rf drive~\cite{Meinelt_2024}:
\begin{equation}
    \label{Eqn:power}
    P \propto V_{rf}^2\Omega_{rf}^2 .
\end{equation}
As seen in Table~\ref{Table:example}, the cross-rf (gnd-surface) trap will need orders-of-magnitude less power with respect to a surface trap due to reduced drive amplitude and frequency. This further highlights the need for high harmonicity to enable operation at $q\sim1$. 

\begin{table}
 \centering
\caption{\label{Table:example} Experimental values of the investigated FoMs and the heating and power scaling for each trap geometry. The errors on the reported values are discussed in the text and are limited by the resolution of the simulations which limits the estimation of $d$.}
\begin{tabular}{c|ccccccc}
Geometry & $d$ & $k$ & $q$ & $\omega_\text{rad} /2\pi$  & $V_\text{rf}$ & $\Omega_\text{rf} / 2\pi$ & $\bar{P}$  \\ 
& [\textmu m] & & &[MHz] & [kV] & [MHz] & (norm)\\ \hline 
surface & 90 & 0.210 & 0.250 & 10  & 15 & 110 & 1  \\
gnd-surface & 46 & 0.280 & 0.333 & 10  & 4.0 & 85 & $ 4.0 \times 10^{-2}$  \\
cross-rf & 60 & 0.697 & 0.905 & 10 &  0.42 & 31 & $ 6.1 \times 10^{-5}$ 
\end{tabular}
\end{table}


\section{Practicality}
\label{Sec:Fab}
\subsection{Fabrication}

Sections~\ref{Sec:Characterization} and \ref{example} suggest that the two multi-wafer designs present a path forward to achieving significantly reduced power dissipation for a given target trap frequency, a key to increasing the number of trap sites in a trapped-ion quantum computer. However, multi-wafer traps complicate fabrication and reduce the overall optical access. Here we consider how the increased complexity of the multi-wafer traps will affect the overall reliability and performance. 

Surface-electrode ion traps can be fabricated using traditional clean-room microfabrication techniques~\cite{Chiaverini_2005}. Simple planar geometries are straightforward to make on a variety of different substrates, and advanced fabrication techniques have been leveraged to create complex surface electrode traps offering specific functionalities such as high optical access through the trap~\cite{Maunz_2016} and junctions between trapping zones~\cite{Amini2010Mar, Burton2023Apr, Moehring2011Jul,Shu2014Jun}. 

Multi-wafer traps are more difficult to fabricate that surface traps, but traps with geometries similar to the cross-rf trap discussed here have been demonstrated with a variety of different techniques. Solutions include manual wafer stacking~\cite{Decaroli_2021,Ragg_2019} and cleanroom-based MEMS techniques for monolithic ion traps~\cite{Wilpers_2012}.  Wafer alignment can be quite precise, with reported angular, lateral, and wafer separation errors of less than 0.05$^{\circ}$, 2\,\textmu m, and 5\,\textmu m, respectively for a 2\,cm $\times$ 2\, cm chip~\cite{Ragg_2019}. 

The gnd-surface trap is significantly simpler than the cross-rf multi-wafer trap and can take advantage of all the microfabrication techniques developed for high performance surface traps. The ground plane is unpatterned and thus lateral misalignment has no impact on the trap parameters. Errors in the trap height will change the ion position but will not distort the potential. Thus, though it provides more modest gains with respect to a surface trap the simplicity may improve repeatability and reliability

\subsection{Optical Access}
The main focus of this manuscript has been the effect of trap geometry on the radial trapping potential. However, a truly scalable architecture must also include optical control including fast control of the spectral, temporal, and spatial degrees of freedom across the visible spectrum~\cite{Park2024Sep}. In general, current experiments take advantage of the large optical access of well-engineered surface traps to excite ions with lasers parallel to the trap and collect fluorescence from the top hemisphere, which has unimpeded optical access. However, as architectures begin to tile trapping regions in a 2D plane~\cite{Delaney_2024,Moses_2023} this will not be sufficient. Instead, the optical control beams will likely need to come from below or above the trap. In surface traps, this has been achieved by integrating slits under the ions~\cite{Maunz_2016} or with integrated photonics and grating couplers~\cite{Mehta2016Dec,Mehta2020Oct,Niffenegger2020Oct,Shirao2022Jun}. 

In previous demonstrations of multi-wafer traps similar to our cross-rf design, optical access was achieved through the opening between the rf and ground electrodes both perpendicular and parallel to the wafers~\cite{Ragg_2019,Decaroli_2021, Wilpers_2012, Hensinger2006Jan,Blakestad2009Apr,Blakestad2011Sep}. In this case, the electrode-electrode separation will define the achievable numerical aperture for both excitation and fluorescence collection. The gnd-surface trap design does not naturally provide the same separation between electrodes for optical access. However, we propose two possible solutions. First, the ground plane could be made from a transparent wafer (e.g. borosilicate) with a grounded coating of a transparent conductive material such as indium-tin-oxide. This would enable all of the benefits laid out in the previous sections without a large change the optical access with respect to the surface trap. However, there will likely still be small reflections from each interface and care will have to be taken to reduce the effect of this stray light. Second, openings in the ground plane could be engineered for collection and excitation, similar to previous high optical access surface traps~\cite{Maunz_2016}. While this would change the electrical trap properties, co-optimization of electronic and optical propoerties is an opportunity for future work.

\section{Conclusion and Outlook}

The geometry of an ion trap's electrodes defines the shape of the radial trapping potential which in turn affects the operation of the trapped-ion device. Here we present a study of how overall design choices affect the experimental performance and ultimately the scalability of a trapped-ion device. We compare three generic trap geometries: (1) a surface trap (2) a miniaturized 3D linear Paul trap and (3) a multi-wafer design comprising a surface trap with the same dimensions as in (1) but with a ground plane above. We do not perform a complete optimization of any of the three geometries but rather consider how the wafer separation affects the trapping parameters and scalability.

We find that the two multi-wafer trap designs provide improved harmonicity, trap depth, and trap frequency. In particular, we find that the improved harmonicity in the multi-wafer traps is particularly important for increasing the maximum trap frequency as it increases the experimentally achievable stability parameter~\cite{Ghosh_1995}. This is particularly evident when considering the maximum achievable trap frequency for a given rf drive voltage or conversely the expected power dissipation for a target trap frequency. In both cases the cross-rf trap provides orders-of-magnitude improvement over a surface trap design and the gnd-surface trap provides more modest, but still significant improvement. While it does not provide the same confinement as a miniturized 3D linear Paul trap, the gnd-surface trap presents a promising balance between improved confinement and manufacturability, making it a viable candidate for scalable trapped-ion quantum technologies. Future work should be done to co-optimize the electrodes with the necessary optical access for qubit control and measurement.
As the demand for quantum technologies grows, optimizing ion trap architectures will be essential for the realization of practical, large-scale quantum processors.

\section{Acknowledgments}
We acknowledge helpful conversations with Christian Pluchar and support from NSF award ECCS-2240291.

\bibliography{forArXiv_v2}

\begin{thebibliography}{68}%
\makeatletter
\providecommand \@ifxundefined [1]{%
 \@ifx{#1\undefined}
}%
\providecommand \@ifnum [1]{%
 \ifnum #1\expandafter \@firstoftwo
 \else \expandafter \@secondoftwo
 \fi
}%
\providecommand \@ifx [1]{%
 \ifx #1\expandafter \@firstoftwo
 \else \expandafter \@secondoftwo
 \fi
}%
\providecommand \natexlab [1]{#1}%
\providecommand \enquote  [1]{``#1''}%
\providecommand \bibnamefont  [1]{#1}%
\providecommand \bibfnamefont [1]{#1}%
\providecommand \citenamefont [1]{#1}%
\providecommand \href@noop [0]{\@secondoftwo}%
\providecommand \href [0]{\begingroup \@sanitize@url \@href}%
\providecommand \@href[1]{\@@startlink{#1}\@@href}%
\providecommand \@@href[1]{\endgroup#1\@@endlink}%
\providecommand \@sanitize@url [0]{\catcode `\\12\catcode `\$12\catcode
  `\&12\catcode `\#12\catcode `\^12\catcode `\_12\catcode `\%12\relax}%
\providecommand \@@startlink[1]{}%
\providecommand \@@endlink[0]{}%
\providecommand \url  [0]{\begingroup\@sanitize@url \@url }%
\providecommand \@url [1]{\endgroup\@href {#1}{\urlprefix }}%
\providecommand \urlprefix  [0]{URL }%
\providecommand \Eprint [0]{\href }%
\providecommand \doibase [0]{http://dx.doi.org/}%
\providecommand \selectlanguage [0]{\@gobble}%
\providecommand \bibinfo  [0]{\@secondoftwo}%
\providecommand \bibfield  [0]{\@secondoftwo}%
\providecommand \translation [1]{[#1]}%
\providecommand \BibitemOpen [0]{}%
\providecommand \bibitemStop [0]{}%
\providecommand \bibitemNoStop [0]{.\EOS\space}%
\providecommand \EOS [0]{\spacefactor3000\relax}%
\providecommand \BibitemShut  [1]{\csname bibitem#1\endcsname}%
\let\auto@bib@innerbib\@empty
\bibitem [{\citenamefont {Ludlow}\ \emph {et~al.}(2015)\citenamefont {Ludlow},
  \citenamefont {Boyd}, \citenamefont {Ye}, \citenamefont {Peik},\ and\
  \citenamefont {Schmidt}}]{Ludlow_2015}%
  \BibitemOpen
  \bibfield  {author} {\bibinfo {author} {\bibfnamefont {A.~D.}\ \bibnamefont
  {Ludlow}}, \bibinfo {author} {\bibfnamefont {M.~M.}\ \bibnamefont {Boyd}},
  \bibinfo {author} {\bibfnamefont {J.}~\bibnamefont {Ye}}, \bibinfo {author}
  {\bibfnamefont {E.}~\bibnamefont {Peik}}, \ and\ \bibinfo {author}
  {\bibfnamefont {P.~O.}\ \bibnamefont {Schmidt}},\ }\href {\doibase
  10.1103/RevModPhys.87.637} {\bibfield  {journal} {\bibinfo  {journal} {Rev.
  Mod. Phys.}\ }\textbf {\bibinfo {volume} {87}},\ \bibinfo {pages} {637}
  (\bibinfo {year} {2015})}\BibitemShut {NoStop}%
\bibitem [{\citenamefont {Gilmore}\ \emph {et~al.}(2021)\citenamefont
  {Gilmore}, \citenamefont {Affolter}, \citenamefont {Lewis-Swan},
  \citenamefont {Barberena}, \citenamefont {Jordan}, \citenamefont {Rey},\ and\
  \citenamefont {Bollinger}}]{Gilmore2021Aug}%
  \BibitemOpen
  \bibfield  {author} {\bibinfo {author} {\bibfnamefont {K.~A.}\ \bibnamefont
  {Gilmore}}, \bibinfo {author} {\bibfnamefont {M.}~\bibnamefont {Affolter}},
  \bibinfo {author} {\bibfnamefont {R.~J.}\ \bibnamefont {Lewis-Swan}},
  \bibinfo {author} {\bibfnamefont {D.}~\bibnamefont {Barberena}}, \bibinfo
  {author} {\bibfnamefont {E.}~\bibnamefont {Jordan}}, \bibinfo {author}
  {\bibfnamefont {A.~M.}\ \bibnamefont {Rey}}, \ and\ \bibinfo {author}
  {\bibfnamefont {J.~J.}\ \bibnamefont {Bollinger}},\ }\href {\doibase
  10.1126/science.abi5226} {\bibfield  {journal} {\bibinfo  {journal}
  {Science}\ }\textbf {\bibinfo {volume} {373}},\ \bibinfo {pages} {673}
  (\bibinfo {year} {2021})}\BibitemShut {NoStop}%
\bibitem [{\citenamefont {Blatt}\ and\ \citenamefont
  {Roos}(2012)}]{Blatt2012Apr}%
  \BibitemOpen
  \bibfield  {author} {\bibinfo {author} {\bibfnamefont {R.}~\bibnamefont
  {Blatt}}\ and\ \bibinfo {author} {\bibfnamefont {C.~F.}\ \bibnamefont
  {Roos}},\ }\href {\doibase 10.1038/nphys2252} {\bibfield  {journal} {\bibinfo
   {journal} {Nat. Phys.}\ }\textbf {\bibinfo {volume} {8}},\ \bibinfo {pages}
  {277} (\bibinfo {year} {2012})}\BibitemShut {NoStop}%
\bibitem [{\citenamefont {Wineland}\ \emph {et~al.}(2003)\citenamefont
  {Wineland}, \citenamefont {Barrett}, \citenamefont {Britton}, \citenamefont
  {Chiaverini}, \citenamefont {DeMarco}, \citenamefont {Itano}, \citenamefont
  {Jelenković}, \citenamefont {Langer}, \citenamefont {Leibfried},
  \citenamefont {Meyer}, \citenamefont {Rosenband},\ and\ \citenamefont
  {Schätz}}]{Wineland_2003}%
  \BibitemOpen
  \bibfield  {author} {\bibinfo {author} {\bibfnamefont {D.~J.}\ \bibnamefont
  {Wineland}}, \bibinfo {author} {\bibfnamefont {M.}~\bibnamefont {Barrett}},
  \bibinfo {author} {\bibfnamefont {J.}~\bibnamefont {Britton}}, \bibinfo
  {author} {\bibfnamefont {J.}~\bibnamefont {Chiaverini}}, \bibinfo {author}
  {\bibfnamefont {B.}~\bibnamefont {DeMarco}}, \bibinfo {author} {\bibfnamefont
  {W.~M.}\ \bibnamefont {Itano}}, \bibinfo {author} {\bibfnamefont
  {B.}~\bibnamefont {Jelenković}}, \bibinfo {author} {\bibfnamefont
  {C.}~\bibnamefont {Langer}}, \bibinfo {author} {\bibfnamefont
  {D.}~\bibnamefont {Leibfried}}, \bibinfo {author} {\bibfnamefont
  {V.}~\bibnamefont {Meyer}}, \bibinfo {author} {\bibfnamefont
  {T.}~\bibnamefont {Rosenband}}, \ and\ \bibinfo {author} {\bibfnamefont
  {T.}~\bibnamefont {Schätz}},\ }\href {\doibase 10.1098/rsta.2003.1205}
  {\bibfield  {journal} {\bibinfo  {journal} {Philosophical Transactions of the
  Royal Society of London. Series A: Mathematical, Physical and Engineering
  Sciences}\ }\textbf {\bibinfo {volume} {361}},\ \bibinfo {pages}
  {1349–1361} (\bibinfo {year} {2003})}\BibitemShut {NoStop}%
\bibitem [{\citenamefont {Schindler}\ \emph {et~al.}(2013)\citenamefont
  {Schindler}, \citenamefont {Nigg}, \citenamefont {Monz}, \citenamefont
  {Barreiro}, \citenamefont {Martinez}, \citenamefont {Wang}, \citenamefont
  {Quint}, \citenamefont {Brandl}, \citenamefont {Nebendahl}, \citenamefont
  {Roos}, \citenamefont {Chwalla}, \citenamefont {Hennrich},\ and\
  \citenamefont {Blatt}}]{Schindler_2013}%
  \BibitemOpen
  \bibfield  {author} {\bibinfo {author} {\bibfnamefont {P.}~\bibnamefont
  {Schindler}}, \bibinfo {author} {\bibfnamefont {D.}~\bibnamefont {Nigg}},
  \bibinfo {author} {\bibfnamefont {T.}~\bibnamefont {Monz}}, \bibinfo {author}
  {\bibfnamefont {J.~T.}\ \bibnamefont {Barreiro}}, \bibinfo {author}
  {\bibfnamefont {E.}~\bibnamefont {Martinez}}, \bibinfo {author}
  {\bibfnamefont {S.~X.}\ \bibnamefont {Wang}}, \bibinfo {author}
  {\bibfnamefont {S.}~\bibnamefont {Quint}}, \bibinfo {author} {\bibfnamefont
  {M.~F.}\ \bibnamefont {Brandl}}, \bibinfo {author} {\bibfnamefont
  {V.}~\bibnamefont {Nebendahl}}, \bibinfo {author} {\bibfnamefont {C.~F.}\
  \bibnamefont {Roos}}, \bibinfo {author} {\bibfnamefont {M.}~\bibnamefont
  {Chwalla}}, \bibinfo {author} {\bibfnamefont {M.}~\bibnamefont {Hennrich}}, \
  and\ \bibinfo {author} {\bibfnamefont {R.}~\bibnamefont {Blatt}},\ }\href
  {\doibase 10.1088/1367-2630/15/12/123012} {\bibfield  {journal} {\bibinfo
  {journal} {New Journal of Physics}\ }\textbf {\bibinfo {volume} {15}},\
  \bibinfo {pages} {123012} (\bibinfo {year} {2013})}\BibitemShut {NoStop}%
\bibitem [{\citenamefont {An}\ \emph {et~al.}(2022)\citenamefont {An},
  \citenamefont {Ransford}, \citenamefont {Schaffer}, \citenamefont {Sletten},
  \citenamefont {Gaebler}, \citenamefont {Hostetter},\ and\ \citenamefont
  {Vittorini}}]{An_2022}%
  \BibitemOpen
  \bibfield  {author} {\bibinfo {author} {\bibfnamefont {F.~A.}\ \bibnamefont
  {An}}, \bibinfo {author} {\bibfnamefont {A.}~\bibnamefont {Ransford}},
  \bibinfo {author} {\bibfnamefont {A.}~\bibnamefont {Schaffer}}, \bibinfo
  {author} {\bibfnamefont {L.~R.}\ \bibnamefont {Sletten}}, \bibinfo {author}
  {\bibfnamefont {J.}~\bibnamefont {Gaebler}}, \bibinfo {author} {\bibfnamefont
  {J.}~\bibnamefont {Hostetter}}, \ and\ \bibinfo {author} {\bibfnamefont
  {G.}~\bibnamefont {Vittorini}},\ }\href {\doibase
  10.1103/physrevlett.129.130501} {\bibfield  {journal} {\bibinfo  {journal}
  {Physical Review Letters}\ }\textbf {\bibinfo {volume} {129}} (\bibinfo
  {year} {2022}),\ 10.1103/physrevlett.129.130501}\BibitemShut {NoStop}%
\bibitem [{\citenamefont {Ballance}\ \emph {et~al.}(2016)\citenamefont
  {Ballance}, \citenamefont {Harty}, \citenamefont {Linke}, \citenamefont
  {Sepiol},\ and\ \citenamefont {Lucas}}]{Ballance_2016}%
  \BibitemOpen
  \bibfield  {author} {\bibinfo {author} {\bibfnamefont {C.~J.}\ \bibnamefont
  {Ballance}}, \bibinfo {author} {\bibfnamefont {T.~P.}\ \bibnamefont {Harty}},
  \bibinfo {author} {\bibfnamefont {N.~M.}\ \bibnamefont {Linke}}, \bibinfo
  {author} {\bibfnamefont {M.~A.}\ \bibnamefont {Sepiol}}, \ and\ \bibinfo
  {author} {\bibfnamefont {D.~M.}\ \bibnamefont {Lucas}},\ }\href {\doibase
  10.1103/PhysRevLett.117.060504} {\bibfield  {journal} {\bibinfo  {journal}
  {Phys. Rev. Lett.}\ }\textbf {\bibinfo {volume} {117}},\ \bibinfo {pages}
  {060504} (\bibinfo {year} {2016})}\BibitemShut {NoStop}%
\bibitem [{\citenamefont {Clark}\ \emph {et~al.}(2021)\citenamefont {Clark},
  \citenamefont {Tinkey}, \citenamefont {Sawyer}, \citenamefont {Meier},
  \citenamefont {Burkhardt}, \citenamefont {Seck}, \citenamefont {Shappert},
  \citenamefont {Guise}, \citenamefont {Volin}, \citenamefont {Fallek},
  \citenamefont {Hayden}, \citenamefont {Rellergert},\ and\ \citenamefont
  {Brown}}]{Clark_2021}%
  \BibitemOpen
  \bibfield  {author} {\bibinfo {author} {\bibfnamefont {C.~R.}\ \bibnamefont
  {Clark}}, \bibinfo {author} {\bibfnamefont {H.~N.}\ \bibnamefont {Tinkey}},
  \bibinfo {author} {\bibfnamefont {B.~C.}\ \bibnamefont {Sawyer}}, \bibinfo
  {author} {\bibfnamefont {A.~M.}\ \bibnamefont {Meier}}, \bibinfo {author}
  {\bibfnamefont {K.~A.}\ \bibnamefont {Burkhardt}}, \bibinfo {author}
  {\bibfnamefont {C.~M.}\ \bibnamefont {Seck}}, \bibinfo {author}
  {\bibfnamefont {C.~M.}\ \bibnamefont {Shappert}}, \bibinfo {author}
  {\bibfnamefont {N.~D.}\ \bibnamefont {Guise}}, \bibinfo {author}
  {\bibfnamefont {C.~E.}\ \bibnamefont {Volin}}, \bibinfo {author}
  {\bibfnamefont {S.~D.}\ \bibnamefont {Fallek}}, \bibinfo {author}
  {\bibfnamefont {H.~T.}\ \bibnamefont {Hayden}}, \bibinfo {author}
  {\bibfnamefont {W.~G.}\ \bibnamefont {Rellergert}}, \ and\ \bibinfo {author}
  {\bibfnamefont {K.~R.}\ \bibnamefont {Brown}},\ }\href@noop {} {\bibfield
  {journal} {\bibinfo  {journal} {Phys. Rev. Lett.}\ }\textbf {\bibinfo
  {volume} {127}},\ \bibinfo {pages} {130505} (\bibinfo {year}
  {2021})}\BibitemShut {NoStop}%
\bibitem [{\citenamefont {Sawyer}\ and\ \citenamefont
  {Brown}(2021)}]{Sawyer_2021}%
  \BibitemOpen
  \bibfield  {author} {\bibinfo {author} {\bibfnamefont {B.~C.}\ \bibnamefont
  {Sawyer}}\ and\ \bibinfo {author} {\bibfnamefont {K.~R.}\ \bibnamefont
  {Brown}},\ }\href@noop {} {\bibfield  {journal} {\bibinfo  {journal} {Phys.
  Rev. A}\ }\textbf {\bibinfo {volume} {103}},\ \bibinfo {pages} {022427}
  (\bibinfo {year} {2021})}\BibitemShut {NoStop}%
\bibitem [{\citenamefont {Ospelkaus}\ \emph {et~al.}(2011)\citenamefont
  {Ospelkaus}, \citenamefont {Warring}, \citenamefont {Colombe}, \citenamefont
  {Brown}, \citenamefont {Amini}, \citenamefont {Leibfried},\ and\
  \citenamefont {Wineland}}]{Ospelkaus_2011}%
  \BibitemOpen
  \bibfield  {author} {\bibinfo {author} {\bibfnamefont {C.}~\bibnamefont
  {Ospelkaus}}, \bibinfo {author} {\bibfnamefont {U.}~\bibnamefont {Warring}},
  \bibinfo {author} {\bibfnamefont {Y.}~\bibnamefont {Colombe}}, \bibinfo
  {author} {\bibfnamefont {K.~R.}\ \bibnamefont {Brown}}, \bibinfo {author}
  {\bibfnamefont {J.~M.}\ \bibnamefont {Amini}}, \bibinfo {author}
  {\bibfnamefont {D.}~\bibnamefont {Leibfried}}, \ and\ \bibinfo {author}
  {\bibfnamefont {D.~J.}\ \bibnamefont {Wineland}},\ }\href@noop {} {\bibfield
  {journal} {\bibinfo  {journal} {Nature}\ }\textbf {\bibinfo {volume} {476}},\
  \bibinfo {pages} {181–184} (\bibinfo {year} {2011})}\BibitemShut {NoStop}%
\bibitem [{\citenamefont {Weber}\ \emph {et~al.}(2024)\citenamefont {Weber},
  \citenamefont {Gely}, \citenamefont {Hanley}, \citenamefont {Harty},
  \citenamefont {Leu}, \citenamefont {Löschnauer}, \citenamefont {Nadlinger},\
  and\ \citenamefont {Lucas}}]{Weber_2024}%
  \BibitemOpen
  \bibfield  {author} {\bibinfo {author} {\bibfnamefont {M.~A.}\ \bibnamefont
  {Weber}}, \bibinfo {author} {\bibfnamefont {M.~F.}\ \bibnamefont {Gely}},
  \bibinfo {author} {\bibfnamefont {R.~K.}\ \bibnamefont {Hanley}}, \bibinfo
  {author} {\bibfnamefont {T.~P.}\ \bibnamefont {Harty}}, \bibinfo {author}
  {\bibfnamefont {A.~D.}\ \bibnamefont {Leu}}, \bibinfo {author} {\bibfnamefont
  {C.~M.}\ \bibnamefont {Löschnauer}}, \bibinfo {author} {\bibfnamefont
  {D.~P.}\ \bibnamefont {Nadlinger}}, \ and\ \bibinfo {author} {\bibfnamefont
  {D.~M.}\ \bibnamefont {Lucas}},\ }\href@noop {} {\bibfield  {journal}
  {\bibinfo  {journal} {Physical Review A}\ }\textbf {\bibinfo {volume} {110}}
  (\bibinfo {year} {2024})}\BibitemShut {NoStop}%
\bibitem [{\citenamefont {L\"oschnauer}\ \emph {et~al.}(2024)\citenamefont
  {L\"oschnauer} \emph {et~al.}}]{Loschnauer_2024}%
  \BibitemOpen
  \bibfield  {author} {\bibinfo {author} {\bibfnamefont {C.~M.}\ \bibnamefont
  {L\"oschnauer}} \emph {et~al.},\ }\href@noop {} {\bibfield  {journal}
  {\bibinfo  {journal} {arXiv}\ } (\bibinfo {year} {2024})},\ \Eprint
  {http://arxiv.org/abs/2407.07694} {2407.07694 [quant-ph]} \BibitemShut
  {NoStop}%
\bibitem [{\citenamefont {Paul}(1990)}]{Paul_1990}%
  \BibitemOpen
  \bibfield  {author} {\bibinfo {author} {\bibfnamefont {W.}~\bibnamefont
  {Paul}},\ }\href@noop {} {\bibfield  {journal} {\bibinfo  {journal} {Rev.
  Mod. Phys.}\ }\textbf {\bibinfo {volume} {62}},\ \bibinfo {pages} {531}
  (\bibinfo {year} {1990})}\BibitemShut {NoStop}%
\bibitem [{\citenamefont {Chiaverini}\ \emph {et~al.}(2005)\citenamefont
  {Chiaverini}, \citenamefont {Blakestad}, \citenamefont {Britton},
  \citenamefont {Jost}, \citenamefont {Langer}, \citenamefont {Leibfried},
  \citenamefont {Ozeri},\ and\ \citenamefont {Wineland}}]{Chiaverini_2005}%
  \BibitemOpen
  \bibfield  {author} {\bibinfo {author} {\bibfnamefont {J.}~\bibnamefont
  {Chiaverini}}, \bibinfo {author} {\bibfnamefont {R.~B.}\ \bibnamefont
  {Blakestad}}, \bibinfo {author} {\bibfnamefont {J.}~\bibnamefont {Britton}},
  \bibinfo {author} {\bibfnamefont {J.~D.}\ \bibnamefont {Jost}}, \bibinfo
  {author} {\bibfnamefont {C.}~\bibnamefont {Langer}}, \bibinfo {author}
  {\bibfnamefont {D.}~\bibnamefont {Leibfried}}, \bibinfo {author}
  {\bibfnamefont {R.}~\bibnamefont {Ozeri}}, \ and\ \bibinfo {author}
  {\bibfnamefont {D.~J.}\ \bibnamefont {Wineland}},\ }\href@noop {} {\bibfield
  {journal} {\bibinfo  {journal} {Quantum Information and Computation}\
  }\textbf {\bibinfo {volume} {5}},\ \bibinfo {pages} {419} (\bibinfo {year}
  {2005})}\BibitemShut {NoStop}%
\bibitem [{\citenamefont {Zhang}\ \emph {et~al.}(2017)\citenamefont {Zhang},
  \citenamefont {Pagano}, \citenamefont {Hess}, \citenamefont {Kyprianidis},
  \citenamefont {Becker}, \citenamefont {Kaplan}, \citenamefont {Gorshkov},
  \citenamefont {Gong},\ and\ \citenamefont {Monroe}}]{Zhang_2017}%
  \BibitemOpen
  \bibfield  {author} {\bibinfo {author} {\bibfnamefont {J.}~\bibnamefont
  {Zhang}}, \bibinfo {author} {\bibfnamefont {G.}~\bibnamefont {Pagano}},
  \bibinfo {author} {\bibfnamefont {P.~W.}\ \bibnamefont {Hess}}, \bibinfo
  {author} {\bibfnamefont {A.}~\bibnamefont {Kyprianidis}}, \bibinfo {author}
  {\bibfnamefont {P.}~\bibnamefont {Becker}}, \bibinfo {author} {\bibfnamefont
  {H.}~\bibnamefont {Kaplan}}, \bibinfo {author} {\bibfnamefont {A.~V.}\
  \bibnamefont {Gorshkov}}, \bibinfo {author} {\bibfnamefont {Z.-X.}\
  \bibnamefont {Gong}}, \ and\ \bibinfo {author} {\bibfnamefont
  {C.}~\bibnamefont {Monroe}},\ }\href {\doibase 10.1038/nature24654}
  {\bibfield  {journal} {\bibinfo  {journal} {Nature}\ }\textbf {\bibinfo
  {volume} {551}},\ \bibinfo {pages} {601–604} (\bibinfo {year}
  {2017})}\BibitemShut {NoStop}%
\bibitem [{\citenamefont {Pino}\ \emph {et~al.}(2021)\citenamefont {Pino},
  \citenamefont {Dreiling}, \citenamefont {Figgatt}, \citenamefont {Gaebler},
  \citenamefont {Moses}, \citenamefont {Allman}, \citenamefont {Baldwin},
  \citenamefont {Foss-Feig}, \citenamefont {Hayes}, \citenamefont {Mayer},
  \citenamefont {Ryan-Anderson},\ and\ \citenamefont {Neyenhuis}}]{Pino_2021}%
  \BibitemOpen
  \bibfield  {author} {\bibinfo {author} {\bibfnamefont {J.~M.}\ \bibnamefont
  {Pino}}, \bibinfo {author} {\bibfnamefont {J.~M.}\ \bibnamefont {Dreiling}},
  \bibinfo {author} {\bibfnamefont {C.}~\bibnamefont {Figgatt}}, \bibinfo
  {author} {\bibfnamefont {J.~P.}\ \bibnamefont {Gaebler}}, \bibinfo {author}
  {\bibfnamefont {S.~A.}\ \bibnamefont {Moses}}, \bibinfo {author}
  {\bibfnamefont {M.~S.}\ \bibnamefont {Allman}}, \bibinfo {author}
  {\bibfnamefont {C.~H.}\ \bibnamefont {Baldwin}}, \bibinfo {author}
  {\bibfnamefont {M.}~\bibnamefont {Foss-Feig}}, \bibinfo {author}
  {\bibfnamefont {D.}~\bibnamefont {Hayes}}, \bibinfo {author} {\bibfnamefont
  {K.}~\bibnamefont {Mayer}}, \bibinfo {author} {\bibfnamefont
  {C.}~\bibnamefont {Ryan-Anderson}}, \ and\ \bibinfo {author} {\bibfnamefont
  {B.}~\bibnamefont {Neyenhuis}},\ }\href@noop {} {\bibfield  {journal}
  {\bibinfo  {journal} {Nature}\ }\textbf {\bibinfo {volume} {592}},\ \bibinfo
  {pages} {209–213} (\bibinfo {year} {2021})}\BibitemShut {NoStop}%
\bibitem [{\citenamefont {Pogorelov}\ \emph {et~al.}(2021)\citenamefont
  {Pogorelov}, \citenamefont {Feldker}, \citenamefont {Marciniak},
  \citenamefont {Postler}, \citenamefont {Jacob}, \citenamefont
  {Krieglsteiner}, \citenamefont {Podlesnic}, \citenamefont {Meth},
  \citenamefont {Negnevitsky}, \citenamefont {Stadler}, \citenamefont
  {H\"ofer}, \citenamefont {W\"achter}, \citenamefont {Lakhmanskiy},
  \citenamefont {Blatt}, \citenamefont {Schindler},\ and\ \citenamefont
  {Monz}}]{Pogorelov_2021}%
  \BibitemOpen
  \bibfield  {author} {\bibinfo {author} {\bibfnamefont {I.}~\bibnamefont
  {Pogorelov}}, \bibinfo {author} {\bibfnamefont {T.}~\bibnamefont {Feldker}},
  \bibinfo {author} {\bibfnamefont {C.~D.}\ \bibnamefont {Marciniak}}, \bibinfo
  {author} {\bibfnamefont {L.}~\bibnamefont {Postler}}, \bibinfo {author}
  {\bibfnamefont {G.}~\bibnamefont {Jacob}}, \bibinfo {author} {\bibfnamefont
  {O.}~\bibnamefont {Krieglsteiner}}, \bibinfo {author} {\bibfnamefont
  {V.}~\bibnamefont {Podlesnic}}, \bibinfo {author} {\bibfnamefont
  {M.}~\bibnamefont {Meth}}, \bibinfo {author} {\bibfnamefont {V.}~\bibnamefont
  {Negnevitsky}}, \bibinfo {author} {\bibfnamefont {M.}~\bibnamefont
  {Stadler}}, \bibinfo {author} {\bibfnamefont {B.}~\bibnamefont {H\"ofer}},
  \bibinfo {author} {\bibfnamefont {C.}~\bibnamefont {W\"achter}}, \bibinfo
  {author} {\bibfnamefont {K.}~\bibnamefont {Lakhmanskiy}}, \bibinfo {author}
  {\bibfnamefont {R.}~\bibnamefont {Blatt}}, \bibinfo {author} {\bibfnamefont
  {P.}~\bibnamefont {Schindler}}, \ and\ \bibinfo {author} {\bibfnamefont
  {T.}~\bibnamefont {Monz}},\ }\href {\doibase 10.1103/PRXQuantum.2.020343}
  {\bibfield  {journal} {\bibinfo  {journal} {PRX Quantum}\ }\textbf {\bibinfo
  {volume} {2}},\ \bibinfo {pages} {020343} (\bibinfo {year}
  {2021})}\BibitemShut {NoStop}%
\bibitem [{\citenamefont {Moses}\ \emph {et~al.}(2023)\citenamefont {Moses},
  \citenamefont {Baldwin}, \citenamefont {Allman}, \citenamefont {Ancona},
  \citenamefont {Ascarrunz}, \citenamefont {Barnes}, \citenamefont
  {Bartolotta}, \citenamefont {Bjork}, \citenamefont {Blanchard}, \citenamefont
  {Bohn}, \citenamefont {Bohnet}, \citenamefont {Brown}, \citenamefont
  {Burdick}, \citenamefont {Burton}, \citenamefont {Campbell}, \citenamefont
  {Campora}, \citenamefont {Carron}, \citenamefont {Chambers}, \citenamefont
  {Chan}, \citenamefont {Chen}, \citenamefont {Chernoguzov}, \citenamefont
  {Chertkov}, \citenamefont {Colina}, \citenamefont {Curtis}, \citenamefont
  {Daniel}, \citenamefont {DeCross}, \citenamefont {Deen}, \citenamefont
  {Delaney}, \citenamefont {Dreiling}, \citenamefont {Ertsgaard}, \citenamefont
  {Esposito}, \citenamefont {Estey}, \citenamefont {Fabrikant}, \citenamefont
  {Figgatt}, \citenamefont {Foltz}, \citenamefont {Foss-Feig}, \citenamefont
  {Francois}, \citenamefont {Gaebler}, \citenamefont {Gatterman}, \citenamefont
  {Gilbreth}, \citenamefont {Giles}, \citenamefont {Glynn}, \citenamefont
  {Hall}, \citenamefont {Hankin}, \citenamefont {Hansen}, \citenamefont
  {Hayes}, \citenamefont {Higashi}, \citenamefont {Hoffman}, \citenamefont
  {Horning}, \citenamefont {Hout}, \citenamefont {Jacobs}, \citenamefont
  {Johansen}, \citenamefont {Jones}, \citenamefont {Karcz}, \citenamefont
  {Klein}, \citenamefont {Lauria}, \citenamefont {Lee}, \citenamefont {Liefer},
  \citenamefont {Lu}, \citenamefont {Lucchetti}, \citenamefont {Lytle},
  \citenamefont {Malm}, \citenamefont {Matheny}, \citenamefont {Mathewson},
  \citenamefont {Mayer}, \citenamefont {Miller}, \citenamefont {Mills},
  \citenamefont {Neyenhuis}, \citenamefont {Nugent}, \citenamefont {Olson},
  \citenamefont {Parks}, \citenamefont {Price}, \citenamefont {Price},
  \citenamefont {Pugh}, \citenamefont {Ransford}, \citenamefont {Reed},
  \citenamefont {Roman}, \citenamefont {Rowe}, \citenamefont {Ryan-Anderson},
  \citenamefont {Sanders}, \citenamefont {Sedlacek}, \citenamefont {Shevchuk},
  \citenamefont {Siegfried}, \citenamefont {Skripka}, \citenamefont {Spaun},
  \citenamefont {Sprenkle}, \citenamefont {Stutz}, \citenamefont {Swallows},
  \citenamefont {Tobey}, \citenamefont {Tran}, \citenamefont {Tran},
  \citenamefont {Vogt}, \citenamefont {Volin}, \citenamefont {Walker},
  \citenamefont {Zolot},\ and\ \citenamefont {Pino}}]{Moses_2023}%
  \BibitemOpen
  \bibfield  {author} {\bibinfo {author} {\bibfnamefont {S.}~\bibnamefont
  {Moses}}, \bibinfo {author} {\bibfnamefont {C.}~\bibnamefont {Baldwin}},
  \bibinfo {author} {\bibfnamefont {M.}~\bibnamefont {Allman}}, \bibinfo
  {author} {\bibfnamefont {R.}~\bibnamefont {Ancona}}, \bibinfo {author}
  {\bibfnamefont {L.}~\bibnamefont {Ascarrunz}}, \bibinfo {author}
  {\bibfnamefont {C.}~\bibnamefont {Barnes}}, \bibinfo {author} {\bibfnamefont
  {J.}~\bibnamefont {Bartolotta}}, \bibinfo {author} {\bibfnamefont
  {B.}~\bibnamefont {Bjork}}, \bibinfo {author} {\bibfnamefont
  {P.}~\bibnamefont {Blanchard}}, \bibinfo {author} {\bibfnamefont
  {M.}~\bibnamefont {Bohn}}, \bibinfo {author} {\bibfnamefont {J.}~\bibnamefont
  {Bohnet}}, \bibinfo {author} {\bibfnamefont {N.}~\bibnamefont {Brown}},
  \bibinfo {author} {\bibfnamefont {N.}~\bibnamefont {Burdick}}, \bibinfo
  {author} {\bibfnamefont {W.}~\bibnamefont {Burton}}, \bibinfo {author}
  {\bibfnamefont {S.}~\bibnamefont {Campbell}}, \bibinfo {author}
  {\bibfnamefont {J.}~\bibnamefont {Campora}}, \bibinfo {author} {\bibfnamefont
  {C.}~\bibnamefont {Carron}}, \bibinfo {author} {\bibfnamefont
  {J.}~\bibnamefont {Chambers}}, \bibinfo {author} {\bibfnamefont
  {J.}~\bibnamefont {Chan}}, \bibinfo {author} {\bibfnamefont {Y.}~\bibnamefont
  {Chen}}, \bibinfo {author} {\bibfnamefont {A.}~\bibnamefont {Chernoguzov}},
  \bibinfo {author} {\bibfnamefont {E.}~\bibnamefont {Chertkov}}, \bibinfo
  {author} {\bibfnamefont {J.}~\bibnamefont {Colina}}, \bibinfo {author}
  {\bibfnamefont {J.}~\bibnamefont {Curtis}}, \bibinfo {author} {\bibfnamefont
  {R.}~\bibnamefont {Daniel}}, \bibinfo {author} {\bibfnamefont
  {M.}~\bibnamefont {DeCross}}, \bibinfo {author} {\bibfnamefont
  {D.}~\bibnamefont {Deen}}, \bibinfo {author} {\bibfnamefont {C.}~\bibnamefont
  {Delaney}}, \bibinfo {author} {\bibfnamefont {J.}~\bibnamefont {Dreiling}},
  \bibinfo {author} {\bibfnamefont {C.}~\bibnamefont {Ertsgaard}}, \bibinfo
  {author} {\bibfnamefont {J.}~\bibnamefont {Esposito}}, \bibinfo {author}
  {\bibfnamefont {B.}~\bibnamefont {Estey}}, \bibinfo {author} {\bibfnamefont
  {M.}~\bibnamefont {Fabrikant}}, \bibinfo {author} {\bibfnamefont
  {C.}~\bibnamefont {Figgatt}}, \bibinfo {author} {\bibfnamefont
  {C.}~\bibnamefont {Foltz}}, \bibinfo {author} {\bibfnamefont
  {M.}~\bibnamefont {Foss-Feig}}, \bibinfo {author} {\bibfnamefont
  {D.}~\bibnamefont {Francois}}, \bibinfo {author} {\bibfnamefont
  {J.}~\bibnamefont {Gaebler}}, \bibinfo {author} {\bibfnamefont
  {T.}~\bibnamefont {Gatterman}}, \bibinfo {author} {\bibfnamefont
  {C.}~\bibnamefont {Gilbreth}}, \bibinfo {author} {\bibfnamefont
  {J.}~\bibnamefont {Giles}}, \bibinfo {author} {\bibfnamefont
  {E.}~\bibnamefont {Glynn}}, \bibinfo {author} {\bibfnamefont
  {A.}~\bibnamefont {Hall}}, \bibinfo {author} {\bibfnamefont {A.}~\bibnamefont
  {Hankin}}, \bibinfo {author} {\bibfnamefont {A.}~\bibnamefont {Hansen}},
  \bibinfo {author} {\bibfnamefont {D.}~\bibnamefont {Hayes}}, \bibinfo
  {author} {\bibfnamefont {B.}~\bibnamefont {Higashi}}, \bibinfo {author}
  {\bibfnamefont {I.}~\bibnamefont {Hoffman}}, \bibinfo {author} {\bibfnamefont
  {B.}~\bibnamefont {Horning}}, \bibinfo {author} {\bibfnamefont
  {J.}~\bibnamefont {Hout}}, \bibinfo {author} {\bibfnamefont {R.}~\bibnamefont
  {Jacobs}}, \bibinfo {author} {\bibfnamefont {J.}~\bibnamefont {Johansen}},
  \bibinfo {author} {\bibfnamefont {L.}~\bibnamefont {Jones}}, \bibinfo
  {author} {\bibfnamefont {J.}~\bibnamefont {Karcz}}, \bibinfo {author}
  {\bibfnamefont {T.}~\bibnamefont {Klein}}, \bibinfo {author} {\bibfnamefont
  {P.}~\bibnamefont {Lauria}}, \bibinfo {author} {\bibfnamefont
  {P.}~\bibnamefont {Lee}}, \bibinfo {author} {\bibfnamefont {D.}~\bibnamefont
  {Liefer}}, \bibinfo {author} {\bibfnamefont {S.}~\bibnamefont {Lu}}, \bibinfo
  {author} {\bibfnamefont {D.}~\bibnamefont {Lucchetti}}, \bibinfo {author}
  {\bibfnamefont {C.}~\bibnamefont {Lytle}}, \bibinfo {author} {\bibfnamefont
  {A.}~\bibnamefont {Malm}}, \bibinfo {author} {\bibfnamefont {M.}~\bibnamefont
  {Matheny}}, \bibinfo {author} {\bibfnamefont {B.}~\bibnamefont {Mathewson}},
  \bibinfo {author} {\bibfnamefont {K.}~\bibnamefont {Mayer}}, \bibinfo
  {author} {\bibfnamefont {D.}~\bibnamefont {Miller}}, \bibinfo {author}
  {\bibfnamefont {M.}~\bibnamefont {Mills}}, \bibinfo {author} {\bibfnamefont
  {B.}~\bibnamefont {Neyenhuis}}, \bibinfo {author} {\bibfnamefont
  {L.}~\bibnamefont {Nugent}}, \bibinfo {author} {\bibfnamefont
  {S.}~\bibnamefont {Olson}}, \bibinfo {author} {\bibfnamefont
  {J.}~\bibnamefont {Parks}}, \bibinfo {author} {\bibfnamefont
  {G.}~\bibnamefont {Price}}, \bibinfo {author} {\bibfnamefont
  {Z.}~\bibnamefont {Price}}, \bibinfo {author} {\bibfnamefont
  {M.}~\bibnamefont {Pugh}}, \bibinfo {author} {\bibfnamefont {A.}~\bibnamefont
  {Ransford}}, \bibinfo {author} {\bibfnamefont {A.}~\bibnamefont {Reed}},
  \bibinfo {author} {\bibfnamefont {C.}~\bibnamefont {Roman}}, \bibinfo
  {author} {\bibfnamefont {M.}~\bibnamefont {Rowe}}, \bibinfo {author}
  {\bibfnamefont {C.}~\bibnamefont {Ryan-Anderson}}, \bibinfo {author}
  {\bibfnamefont {S.}~\bibnamefont {Sanders}}, \bibinfo {author} {\bibfnamefont
  {J.}~\bibnamefont {Sedlacek}}, \bibinfo {author} {\bibfnamefont
  {P.}~\bibnamefont {Shevchuk}}, \bibinfo {author} {\bibfnamefont
  {P.}~\bibnamefont {Siegfried}}, \bibinfo {author} {\bibfnamefont
  {T.}~\bibnamefont {Skripka}}, \bibinfo {author} {\bibfnamefont
  {B.}~\bibnamefont {Spaun}}, \bibinfo {author} {\bibfnamefont
  {R.}~\bibnamefont {Sprenkle}}, \bibinfo {author} {\bibfnamefont
  {R.}~\bibnamefont {Stutz}}, \bibinfo {author} {\bibfnamefont
  {M.}~\bibnamefont {Swallows}}, \bibinfo {author} {\bibfnamefont
  {R.}~\bibnamefont {Tobey}}, \bibinfo {author} {\bibfnamefont
  {A.}~\bibnamefont {Tran}}, \bibinfo {author} {\bibfnamefont {T.}~\bibnamefont
  {Tran}}, \bibinfo {author} {\bibfnamefont {E.}~\bibnamefont {Vogt}}, \bibinfo
  {author} {\bibfnamefont {C.}~\bibnamefont {Volin}}, \bibinfo {author}
  {\bibfnamefont {J.}~\bibnamefont {Walker}}, \bibinfo {author} {\bibfnamefont
  {A.}~\bibnamefont {Zolot}}, \ and\ \bibinfo {author} {\bibfnamefont
  {J.}~\bibnamefont {Pino}},\ }\href@noop {} {\bibfield  {journal} {\bibinfo
  {journal} {Physical Review X}\ }\textbf {\bibinfo {volume} {13}} (\bibinfo
  {year} {2023})}\BibitemShut {NoStop}%
\bibitem [{\citenamefont {Chen}\ \emph {et~al.}(2024)\citenamefont {Chen},
  \citenamefont {Nielsen}, \citenamefont {Ebert}, \citenamefont {Inlek},
  \citenamefont {Wright}, \citenamefont {Chaplin}, \citenamefont {Maksymov},
  \citenamefont {P{\'{a}}ez}, \citenamefont {Poudel}, \citenamefont {Maunz},\
  and\ \citenamefont {Gamble}}]{Chen_2024}%
  \BibitemOpen
  \bibfield  {author} {\bibinfo {author} {\bibfnamefont {J.-S.}\ \bibnamefont
  {Chen}}, \bibinfo {author} {\bibfnamefont {E.}~\bibnamefont {Nielsen}},
  \bibinfo {author} {\bibfnamefont {M.}~\bibnamefont {Ebert}}, \bibinfo
  {author} {\bibfnamefont {V.}~\bibnamefont {Inlek}}, \bibinfo {author}
  {\bibfnamefont {K.}~\bibnamefont {Wright}}, \bibinfo {author} {\bibfnamefont
  {V.}~\bibnamefont {Chaplin}}, \bibinfo {author} {\bibfnamefont
  {A.}~\bibnamefont {Maksymov}}, \bibinfo {author} {\bibfnamefont
  {E.}~\bibnamefont {P{\'{a}}ez}}, \bibinfo {author} {\bibfnamefont
  {A.}~\bibnamefont {Poudel}}, \bibinfo {author} {\bibfnamefont
  {P.}~\bibnamefont {Maunz}}, \ and\ \bibinfo {author} {\bibfnamefont
  {J.}~\bibnamefont {Gamble}},\ }\href {\doibase 10.22331/q-2024-11-07-1516}
  {\bibfield  {journal} {\bibinfo  {journal} {{Quantum}}\ }\textbf {\bibinfo
  {volume} {8}},\ \bibinfo {pages} {1516} (\bibinfo {year} {2024})}\BibitemShut
  {NoStop}%
\bibitem [{\citenamefont {Postler}\ \emph {et~al.}(2024)\citenamefont
  {Postler}, \citenamefont {Butt}, \citenamefont {Pogorelov}, \citenamefont
  {Marciniak}, \citenamefont {Heu\ss{}en}, \citenamefont {Blatt}, \citenamefont
  {Schindler}, \citenamefont {Rispler}, \citenamefont {M\"uller},\ and\
  \citenamefont {Monz}}]{Postler_2024}%
  \BibitemOpen
  \bibfield  {author} {\bibinfo {author} {\bibfnamefont {L.}~\bibnamefont
  {Postler}}, \bibinfo {author} {\bibfnamefont {F.}~\bibnamefont {Butt}},
  \bibinfo {author} {\bibfnamefont {I.}~\bibnamefont {Pogorelov}}, \bibinfo
  {author} {\bibfnamefont {C.~D.}\ \bibnamefont {Marciniak}}, \bibinfo {author}
  {\bibfnamefont {S.}~\bibnamefont {Heu\ss{}en}}, \bibinfo {author}
  {\bibfnamefont {R.}~\bibnamefont {Blatt}}, \bibinfo {author} {\bibfnamefont
  {P.}~\bibnamefont {Schindler}}, \bibinfo {author} {\bibfnamefont
  {M.}~\bibnamefont {Rispler}}, \bibinfo {author} {\bibfnamefont
  {M.}~\bibnamefont {M\"uller}}, \ and\ \bibinfo {author} {\bibfnamefont
  {T.}~\bibnamefont {Monz}},\ }\href {\doibase 10.1103/PRXQuantum.5.030326}
  {\bibfield  {journal} {\bibinfo  {journal} {PRX Quantum}\ }\textbf {\bibinfo
  {volume} {5}},\ \bibinfo {pages} {030326} (\bibinfo {year}
  {2024})}\BibitemShut {NoStop}%
\bibitem [{\citenamefont {Reichardt}\ \emph {et~al.}(2024)\citenamefont
  {Reichardt}, \citenamefont {Aasen}, \citenamefont {Chao}, \citenamefont
  {Chernoguzov}, \citenamefont {van Dam}, \citenamefont {Gaebler},
  \citenamefont {Gresh}, \citenamefont {Lucchetti}, \citenamefont {Mills},
  \citenamefont {Moses}, \citenamefont {Neyenhuis}, \citenamefont {Paetznick},
  \citenamefont {Paz}, \citenamefont {Siegfried}, \citenamefont {da~Silva},
  \citenamefont {Svore}, \citenamefont {Wang},\ and\ \citenamefont
  {Zanner}}]{Reichardt_2024}%
  \BibitemOpen
  \bibfield  {author} {\bibinfo {author} {\bibfnamefont {B.~W.}\ \bibnamefont
  {Reichardt}}, \bibinfo {author} {\bibfnamefont {D.}~\bibnamefont {Aasen}},
  \bibinfo {author} {\bibfnamefont {R.}~\bibnamefont {Chao}}, \bibinfo {author}
  {\bibfnamefont {A.}~\bibnamefont {Chernoguzov}}, \bibinfo {author}
  {\bibfnamefont {W.}~\bibnamefont {van Dam}}, \bibinfo {author} {\bibfnamefont
  {J.~P.}\ \bibnamefont {Gaebler}}, \bibinfo {author} {\bibfnamefont
  {D.}~\bibnamefont {Gresh}}, \bibinfo {author} {\bibfnamefont
  {D.}~\bibnamefont {Lucchetti}}, \bibinfo {author} {\bibfnamefont
  {M.}~\bibnamefont {Mills}}, \bibinfo {author} {\bibfnamefont {S.~A.}\
  \bibnamefont {Moses}}, \bibinfo {author} {\bibfnamefont {B.}~\bibnamefont
  {Neyenhuis}}, \bibinfo {author} {\bibfnamefont {A.}~\bibnamefont
  {Paetznick}}, \bibinfo {author} {\bibfnamefont {A.}~\bibnamefont {Paz}},
  \bibinfo {author} {\bibfnamefont {P.~E.}\ \bibnamefont {Siegfried}}, \bibinfo
  {author} {\bibfnamefont {M.~P.}\ \bibnamefont {da~Silva}}, \bibinfo {author}
  {\bibfnamefont {K.~M.}\ \bibnamefont {Svore}}, \bibinfo {author}
  {\bibfnamefont {Z.}~\bibnamefont {Wang}}, \ and\ \bibinfo {author}
  {\bibfnamefont {M.}~\bibnamefont {Zanner}},\ }\href
  {https://arxiv.org/abs/2409.04628} {\bibfield  {journal} {\bibinfo  {journal}
  {arXiv}\ } (\bibinfo {year} {2024})},\ \Eprint
  {http://arxiv.org/abs/2409.04628} {2409.04628 [quant-ph]} \BibitemShut
  {NoStop}%
\bibitem [{\citenamefont {Zhang}\ \emph {et~al.}(2020)\citenamefont {Zhang},
  \citenamefont {Pokorny}, \citenamefont {Li} \emph {et~al.}}]{Zhang_2020}%
  \BibitemOpen
  \bibfield  {author} {\bibinfo {author} {\bibfnamefont {C.}~\bibnamefont
  {Zhang}}, \bibinfo {author} {\bibfnamefont {F.}~\bibnamefont {Pokorny}},
  \bibinfo {author} {\bibfnamefont {W.}~\bibnamefont {Li}},  \emph {et~al.},\
  }\href {\doibase 10.1038/s41586-020-2152-9} {\bibfield  {journal} {\bibinfo
  {journal} {Nature}\ }\textbf {\bibinfo {volume} {580}},\ \bibinfo {pages}
  {345} (\bibinfo {year} {2020})}\BibitemShut {NoStop}%
\bibitem [{\citenamefont {Saha}\ \emph {et~al.}(2025)\citenamefont {Saha},
  \citenamefont {Shalaev}, \citenamefont {O{'}Reilly}, \citenamefont
  {Goetting}, \citenamefont {Toh}, \citenamefont {Kalakuntla}, \citenamefont
  {Yu},\ and\ \citenamefont {Monroe}}]{Saha2025Mar}%
  \BibitemOpen
  \bibfield  {author} {\bibinfo {author} {\bibfnamefont {S.}~\bibnamefont
  {Saha}}, \bibinfo {author} {\bibfnamefont {M.}~\bibnamefont {Shalaev}},
  \bibinfo {author} {\bibfnamefont {J.}~\bibnamefont {O{'}Reilly}}, \bibinfo
  {author} {\bibfnamefont {I.}~\bibnamefont {Goetting}}, \bibinfo {author}
  {\bibfnamefont {G.}~\bibnamefont {Toh}}, \bibinfo {author} {\bibfnamefont
  {A.}~\bibnamefont {Kalakuntla}}, \bibinfo {author} {\bibfnamefont
  {Y.}~\bibnamefont {Yu}}, \ and\ \bibinfo {author} {\bibfnamefont
  {C.}~\bibnamefont {Monroe}},\ }\href {\doibase 10.1038/s41467-025-57557-4}
  {\bibfield  {journal} {\bibinfo  {journal} {Nat. Commun.}\ }\textbf {\bibinfo
  {volume} {16}},\ \bibinfo {pages} {1} (\bibinfo {year} {2025})}\BibitemShut
  {NoStop}%
\bibitem [{\citenamefont {Saner}\ \emph {et~al.}(2023)\citenamefont {Saner},
  \citenamefont {B\ifmmode \u{a}\else \u{a}\fi{}z\ifmmode~\u{a}\else
  \u{a}\fi{}van}, \citenamefont {Minder}, \citenamefont {Drmota}, \citenamefont
  {Webb}, \citenamefont {Araneda}, \citenamefont {Srinivas}, \citenamefont
  {Lucas},\ and\ \citenamefont {Ballance}}]{Saner_2023}%
  \BibitemOpen
  \bibfield  {author} {\bibinfo {author} {\bibfnamefont {S.}~\bibnamefont
  {Saner}}, \bibinfo {author} {\bibfnamefont {O.}~\bibnamefont {B\ifmmode
  \u{a}\else \u{a}\fi{}z\ifmmode~\u{a}\else \u{a}\fi{}van}}, \bibinfo {author}
  {\bibfnamefont {M.}~\bibnamefont {Minder}}, \bibinfo {author} {\bibfnamefont
  {P.}~\bibnamefont {Drmota}}, \bibinfo {author} {\bibfnamefont {D.~J.}\
  \bibnamefont {Webb}}, \bibinfo {author} {\bibfnamefont {G.}~\bibnamefont
  {Araneda}}, \bibinfo {author} {\bibfnamefont {R.}~\bibnamefont {Srinivas}},
  \bibinfo {author} {\bibfnamefont {D.~M.}\ \bibnamefont {Lucas}}, \ and\
  \bibinfo {author} {\bibfnamefont {C.~J.}\ \bibnamefont {Ballance}},\ }\href
  {\doibase 10.1103/PhysRevLett.131.220601} {\bibfield  {journal} {\bibinfo
  {journal} {Phys. Rev. Lett.}\ }\textbf {\bibinfo {volume} {131}},\ \bibinfo
  {pages} {220601} (\bibinfo {year} {2023})}\BibitemShut {NoStop}%
\bibitem [{\citenamefont {Schäfer}\ \emph {et~al.}(2018)\citenamefont
  {Schäfer}, \citenamefont {Ballance}, \citenamefont {Thirumalai},
  \citenamefont {Stephenson}, \citenamefont {Ballance}, \citenamefont
  {Steane},\ and\ \citenamefont {Lucas}}]{Schafer_2018}%
  \BibitemOpen
  \bibfield  {author} {\bibinfo {author} {\bibfnamefont {V.~M.}\ \bibnamefont
  {Schäfer}}, \bibinfo {author} {\bibfnamefont {C.~J.}\ \bibnamefont
  {Ballance}}, \bibinfo {author} {\bibfnamefont {K.}~\bibnamefont
  {Thirumalai}}, \bibinfo {author} {\bibfnamefont {L.~J.}\ \bibnamefont
  {Stephenson}}, \bibinfo {author} {\bibfnamefont {T.~G.}\ \bibnamefont
  {Ballance}}, \bibinfo {author} {\bibfnamefont {A.~M.}\ \bibnamefont
  {Steane}}, \ and\ \bibinfo {author} {\bibfnamefont {D.~M.}\ \bibnamefont
  {Lucas}},\ }\href {\doibase 10.1038/nature25737} {\bibfield  {journal}
  {\bibinfo  {journal} {Nature}\ }\textbf {\bibinfo {volume} {555}},\ \bibinfo
  {pages} {75–78} (\bibinfo {year} {2018})}\BibitemShut {NoStop}%
\bibitem [{\citenamefont {Steane}\ \emph {et~al.}(2000)\citenamefont {Steane},
  \citenamefont {Roos}, \citenamefont {Stevens}, \citenamefont {Mundt},
  \citenamefont {Leibfried}, \citenamefont {Schmidt-Kaler},\ and\ \citenamefont
  {Blatt}}]{Steane_2000}%
  \BibitemOpen
  \bibfield  {author} {\bibinfo {author} {\bibfnamefont {A.}~\bibnamefont
  {Steane}}, \bibinfo {author} {\bibfnamefont {C.~F.}\ \bibnamefont {Roos}},
  \bibinfo {author} {\bibfnamefont {D.}~\bibnamefont {Stevens}}, \bibinfo
  {author} {\bibfnamefont {A.}~\bibnamefont {Mundt}}, \bibinfo {author}
  {\bibfnamefont {D.}~\bibnamefont {Leibfried}}, \bibinfo {author}
  {\bibfnamefont {F.}~\bibnamefont {Schmidt-Kaler}}, \ and\ \bibinfo {author}
  {\bibfnamefont {R.}~\bibnamefont {Blatt}},\ }\href@noop {} {\bibfield
  {journal} {\bibinfo  {journal} {Phys. Rev. A}\ }\textbf {\bibinfo {volume}
  {62}},\ \bibinfo {pages} {042305} (\bibinfo {year} {2000})}\BibitemShut
  {NoStop}%
\bibitem [{\citenamefont {Brownnutt}\ \emph {et~al.}(2015)\citenamefont
  {Brownnutt}, \citenamefont {Kumph}, \citenamefont {Rabl},\ and\ \citenamefont
  {Blatt}}]{Brownnutt_2015}%
  \BibitemOpen
  \bibfield  {author} {\bibinfo {author} {\bibfnamefont {M.}~\bibnamefont
  {Brownnutt}}, \bibinfo {author} {\bibfnamefont {M.}~\bibnamefont {Kumph}},
  \bibinfo {author} {\bibfnamefont {P.}~\bibnamefont {Rabl}}, \ and\ \bibinfo
  {author} {\bibfnamefont {R.}~\bibnamefont {Blatt}},\ }\href@noop {}
  {\bibfield  {journal} {\bibinfo  {journal} {Rev. Mod. Phys.}\ }\textbf
  {\bibinfo {volume} {87}},\ \bibinfo {pages} {1419} (\bibinfo {year}
  {2015})}\BibitemShut {NoStop}%
\bibitem [{\citenamefont {Ghosh}(1995)}]{Ghosh_1995}%
  \BibitemOpen
  \bibfield  {author} {\bibinfo {author} {\bibfnamefont {P.~K.}\ \bibnamefont
  {Ghosh}},\ }\href@noop {} {\emph {\bibinfo {title} {Ion Traps}}}\ (\bibinfo
  {publisher} {A Clarendon Press Publication},\ \bibinfo {year}
  {1995})\BibitemShut {NoStop}%
\bibitem [{\citenamefont {Kajita}(2022)}]{Kajita_2022}%
  \BibitemOpen
  \bibfield  {author} {\bibinfo {author} {\bibfnamefont {M.}~\bibnamefont
  {Kajita}},\ }\href {\doibase 10.1088/978-0-7503-5472-1} {\emph {\bibinfo
  {title} {Ion Traps}}},\ 2053-2563\ (\bibinfo  {publisher} {IOP Publishing},\
  \bibinfo {year} {2022})\BibitemShut {NoStop}%
\bibitem [{\citenamefont {Pearson}\ \emph {et~al.}(2006)\citenamefont
  {Pearson}, \citenamefont {Leibrandt}, \citenamefont {Bakr}, \citenamefont
  {Mallard}, \citenamefont {Brown},\ and\ \citenamefont
  {Chuang}}]{Pearson_2006}%
  \BibitemOpen
  \bibfield  {author} {\bibinfo {author} {\bibfnamefont {C.~E.}\ \bibnamefont
  {Pearson}}, \bibinfo {author} {\bibfnamefont {D.~R.}\ \bibnamefont
  {Leibrandt}}, \bibinfo {author} {\bibfnamefont {W.~S.}\ \bibnamefont {Bakr}},
  \bibinfo {author} {\bibfnamefont {W.~J.}\ \bibnamefont {Mallard}}, \bibinfo
  {author} {\bibfnamefont {K.~R.}\ \bibnamefont {Brown}}, \ and\ \bibinfo
  {author} {\bibfnamefont {I.~L.}\ \bibnamefont {Chuang}},\ }\href {\doibase
  10.1103/PhysRevA.73.032307} {\bibfield  {journal} {\bibinfo  {journal} {Phys.
  Rev. A}\ }\textbf {\bibinfo {volume} {73}},\ \bibinfo {pages} {032307}
  (\bibinfo {year} {2006})}\BibitemShut {NoStop}%
\bibitem [{\citenamefont {Allcock}\ \emph {et~al.}(2010)\citenamefont
  {Allcock}, \citenamefont {Sherman}, \citenamefont {Stacey}, \citenamefont
  {Burrell}, \citenamefont {Curtis}, \citenamefont {Imreh}, \citenamefont
  {Linke}, \citenamefont {Szwer}, \citenamefont {Webster}, \citenamefont
  {Steane},\ and\ \citenamefont {Lucas}}]{Allcock2010May}%
  \BibitemOpen
  \bibfield  {author} {\bibinfo {author} {\bibfnamefont {D.~T.~C.}\
  \bibnamefont {Allcock}}, \bibinfo {author} {\bibfnamefont {J.~A.}\
  \bibnamefont {Sherman}}, \bibinfo {author} {\bibfnamefont {D.~N.}\
  \bibnamefont {Stacey}}, \bibinfo {author} {\bibfnamefont {A.~H.}\
  \bibnamefont {Burrell}}, \bibinfo {author} {\bibfnamefont {M.~J.}\
  \bibnamefont {Curtis}}, \bibinfo {author} {\bibfnamefont {G.}~\bibnamefont
  {Imreh}}, \bibinfo {author} {\bibfnamefont {N.~M.}\ \bibnamefont {Linke}},
  \bibinfo {author} {\bibfnamefont {D.~J.}\ \bibnamefont {Szwer}}, \bibinfo
  {author} {\bibfnamefont {S.~C.}\ \bibnamefont {Webster}}, \bibinfo {author}
  {\bibfnamefont {A.~M.}\ \bibnamefont {Steane}}, \ and\ \bibinfo {author}
  {\bibfnamefont {D.~M.}\ \bibnamefont {Lucas}},\ }\href {\doibase
  10.1088/1367-2630/12/5/053026} {\bibfield  {journal} {\bibinfo  {journal}
  {New J. Phys.}\ }\textbf {\bibinfo {volume} {12}},\ \bibinfo {pages} {053026}
  (\bibinfo {year} {2010})}\BibitemShut {NoStop}%
\bibitem [{\citenamefont {Hong}\ \emph {et~al.}(2016)\citenamefont {Hong},
  \citenamefont {Lee}, \citenamefont {Cheon}, \citenamefont {Kim},\ and\
  \citenamefont {Cho}}]{Hong_2016}%
  \BibitemOpen
  \bibfield  {author} {\bibinfo {author} {\bibfnamefont {S.}~\bibnamefont
  {Hong}}, \bibinfo {author} {\bibfnamefont {M.}~\bibnamefont {Lee}}, \bibinfo
  {author} {\bibfnamefont {H.}~\bibnamefont {Cheon}}, \bibinfo {author}
  {\bibfnamefont {T.}~\bibnamefont {Kim}}, \ and\ \bibinfo {author}
  {\bibfnamefont {D.-i.}\ \bibnamefont {Cho}},\ }\href {\doibase
  10.3390/s16050616} {\bibfield  {journal} {\bibinfo  {journal} {Sensors}\
  }\textbf {\bibinfo {volume} {16}} (\bibinfo {year} {2016}),\
  10.3390/s16050616}\BibitemShut {NoStop}%
\bibitem [{\citenamefont {Doret}\ \emph {et~al.}(2012)\citenamefont {Doret},
  \citenamefont {Amini}, \citenamefont {Wright}, \citenamefont {Volin},
  \citenamefont {Killian}, \citenamefont {Ozakin}, \citenamefont {Denison},
  \citenamefont {Hayden}, \citenamefont {Pai}, \citenamefont {Slusher},\ and\
  \citenamefont {Harter}}]{Doret2012Jul}%
  \BibitemOpen
  \bibfield  {author} {\bibinfo {author} {\bibfnamefont {S.~C.}\ \bibnamefont
  {Doret}}, \bibinfo {author} {\bibfnamefont {J.~M.}\ \bibnamefont {Amini}},
  \bibinfo {author} {\bibfnamefont {K.}~\bibnamefont {Wright}}, \bibinfo
  {author} {\bibfnamefont {C.}~\bibnamefont {Volin}}, \bibinfo {author}
  {\bibfnamefont {T.}~\bibnamefont {Killian}}, \bibinfo {author} {\bibfnamefont
  {A.}~\bibnamefont {Ozakin}}, \bibinfo {author} {\bibfnamefont
  {D.}~\bibnamefont {Denison}}, \bibinfo {author} {\bibfnamefont
  {H.}~\bibnamefont {Hayden}}, \bibinfo {author} {\bibfnamefont {C.-S.}\
  \bibnamefont {Pai}}, \bibinfo {author} {\bibfnamefont {R.~E.}\ \bibnamefont
  {Slusher}}, \ and\ \bibinfo {author} {\bibfnamefont {A.~W.}\ \bibnamefont
  {Harter}},\ }\href {\doibase 10.1088/1367-2630/14/7/073012} {\bibfield
  {journal} {\bibinfo  {journal} {New J. Phys.}\ }\textbf {\bibinfo {volume}
  {14}},\ \bibinfo {pages} {073012} (\bibinfo {year} {2012})}\BibitemShut
  {NoStop}%
\bibitem [{\citenamefont {Moehring}\ \emph {et~al.}(2011)\citenamefont
  {Moehring}, \citenamefont {Highstrete}, \citenamefont {Stick}, \citenamefont
  {Fortier}, \citenamefont {Haltli}, \citenamefont {Tigges},\ and\
  \citenamefont {Blain}}]{Moehring2011Jul}%
  \BibitemOpen
  \bibfield  {author} {\bibinfo {author} {\bibfnamefont {D.~L.}\ \bibnamefont
  {Moehring}}, \bibinfo {author} {\bibfnamefont {C.}~\bibnamefont
  {Highstrete}}, \bibinfo {author} {\bibfnamefont {D.}~\bibnamefont {Stick}},
  \bibinfo {author} {\bibfnamefont {K.~M.}\ \bibnamefont {Fortier}}, \bibinfo
  {author} {\bibfnamefont {R.}~\bibnamefont {Haltli}}, \bibinfo {author}
  {\bibfnamefont {C.}~\bibnamefont {Tigges}}, \ and\ \bibinfo {author}
  {\bibfnamefont {M.~G.}\ \bibnamefont {Blain}},\ }\href {\doibase
  10.1088/1367-2630/13/7/075018} {\bibfield  {journal} {\bibinfo  {journal}
  {New J. Phys.}\ }\textbf {\bibinfo {volume} {13}},\ \bibinfo {pages} {075018}
  (\bibinfo {year} {2011})}\BibitemShut {NoStop}%
\bibitem [{\citenamefont {Wright}\ \emph {et~al.}(2013)\citenamefont {Wright},
  \citenamefont {Amini}, \citenamefont {Faircloth}, \citenamefont {Volin},
  \citenamefont {Doret}, \citenamefont {Hayden}, \citenamefont {Pai},
  \citenamefont {Landgren}, \citenamefont {Denison}, \citenamefont {Killian},
  \citenamefont {Slusher},\ and\ \citenamefont {Harter}}]{Wright2013Mar}%
  \BibitemOpen
  \bibfield  {author} {\bibinfo {author} {\bibfnamefont {K.}~\bibnamefont
  {Wright}}, \bibinfo {author} {\bibfnamefont {J.~M.}\ \bibnamefont {Amini}},
  \bibinfo {author} {\bibfnamefont {D.~L.}\ \bibnamefont {Faircloth}}, \bibinfo
  {author} {\bibfnamefont {C.}~\bibnamefont {Volin}}, \bibinfo {author}
  {\bibfnamefont {S.~C.}\ \bibnamefont {Doret}}, \bibinfo {author}
  {\bibfnamefont {H.}~\bibnamefont {Hayden}}, \bibinfo {author} {\bibfnamefont
  {C.-S.}\ \bibnamefont {Pai}}, \bibinfo {author} {\bibfnamefont {D.~W.}\
  \bibnamefont {Landgren}}, \bibinfo {author} {\bibfnamefont {D.}~\bibnamefont
  {Denison}}, \bibinfo {author} {\bibfnamefont {T.}~\bibnamefont {Killian}},
  \bibinfo {author} {\bibfnamefont {R.~E.}\ \bibnamefont {Slusher}}, \ and\
  \bibinfo {author} {\bibfnamefont {A.~W.}\ \bibnamefont {Harter}},\ }\href
  {\doibase 10.1088/1367-2630/15/3/033004} {\bibfield  {journal} {\bibinfo
  {journal} {New J. Phys.}\ }\textbf {\bibinfo {volume} {15}},\ \bibinfo
  {pages} {033004} (\bibinfo {year} {2013})}\BibitemShut {NoStop}%
\bibitem [{\citenamefont {Mehta}\ \emph {et~al.}(2016)\citenamefont {Mehta},
  \citenamefont {Bruzewicz}, \citenamefont {McConnell}, \citenamefont {Ram},
  \citenamefont {Sage},\ and\ \citenamefont {Chiaverini}}]{Mehta2016Dec}%
  \BibitemOpen
  \bibfield  {author} {\bibinfo {author} {\bibfnamefont {K.~K.}\ \bibnamefont
  {Mehta}}, \bibinfo {author} {\bibfnamefont {C.~D.}\ \bibnamefont
  {Bruzewicz}}, \bibinfo {author} {\bibfnamefont {R.}~\bibnamefont
  {McConnell}}, \bibinfo {author} {\bibfnamefont {R.~J.}\ \bibnamefont {Ram}},
  \bibinfo {author} {\bibfnamefont {J.~M.}\ \bibnamefont {Sage}}, \ and\
  \bibinfo {author} {\bibfnamefont {J.}~\bibnamefont {Chiaverini}},\ }\href
  {\doibase 10.1038/nnano.2016.139} {\bibfield  {journal} {\bibinfo  {journal}
  {Nat. Nanotechnol.}\ }\textbf {\bibinfo {volume} {11}},\ \bibinfo {pages}
  {1066} (\bibinfo {year} {2016})}\BibitemShut {NoStop}%
\bibitem [{\citenamefont {Maunz}(2016)}]{Maunz_2016}%
  \BibitemOpen
  \bibfield  {author} {\bibinfo {author} {\bibfnamefont {P.~L.~W.}\
  \bibnamefont {Maunz}},\ }\href {\doibase 10.2172/1237003} {\emph {\bibinfo
  {title} {High Optical Access Trap 2.0.}}},\ \bibinfo {type} {Tech. Rep.}\
  (\bibinfo  {institution} {Sandia National Lab. (SNL-NM), Albuquerque, NM
  (United States)},\ \bibinfo {year} {2016})\BibitemShut {NoStop}%
\bibitem [{\citenamefont {Gerasin}\ \emph {et~al.}(2024)\citenamefont
  {Gerasin}, \citenamefont {Zhadnov}, \citenamefont {Kudeyarov}, \citenamefont
  {Khabarova}, \citenamefont {Kolachevsky},\ and\ \citenamefont
  {Semerikov}}]{Gerasin_2024}%
  \BibitemOpen
  \bibfield  {author} {\bibinfo {author} {\bibfnamefont {I.}~\bibnamefont
  {Gerasin}}, \bibinfo {author} {\bibfnamefont {N.}~\bibnamefont {Zhadnov}},
  \bibinfo {author} {\bibfnamefont {K.}~\bibnamefont {Kudeyarov}}, \bibinfo
  {author} {\bibfnamefont {K.}~\bibnamefont {Khabarova}}, \bibinfo {author}
  {\bibfnamefont {N.}~\bibnamefont {Kolachevsky}}, \ and\ \bibinfo {author}
  {\bibfnamefont {I.}~\bibnamefont {Semerikov}},\ }\href {\doibase
  10.3390/quantum6030029} {\bibfield  {journal} {\bibinfo  {journal} {Quantum
  Reports}\ }\textbf {\bibinfo {volume} {6}},\ \bibinfo {pages} {442} (\bibinfo
  {year} {2024})}\BibitemShut {NoStop}%
\bibitem [{\citenamefont {Marquet}\ \emph {et~al.}(2003)\citenamefont
  {Marquet}, \citenamefont {Schmidt-Kaler},\ and\ \citenamefont
  {James}}]{Marquet_2003}%
  \BibitemOpen
  \bibfield  {author} {\bibinfo {author} {\bibfnamefont {C.}~\bibnamefont
  {Marquet}}, \bibinfo {author} {\bibfnamefont {F.}~\bibnamefont
  {Schmidt-Kaler}}, \ and\ \bibinfo {author} {\bibfnamefont {D.}~\bibnamefont
  {James}},\ }\href {\doibase 10.1007/s00340-003-1097-7} {\bibfield  {journal}
  {\bibinfo  {journal} {Applied Physics B: Lasers and Optics}\ }\textbf
  {\bibinfo {volume} {76}},\ \bibinfo {pages} {199–208} (\bibinfo {year}
  {2003})}\BibitemShut {NoStop}%
\bibitem [{\citenamefont {Home}\ \emph {et~al.}(2011)\citenamefont {Home},
  \citenamefont {Hanneke}, \citenamefont {Jost}, \citenamefont {Leibfried},\
  and\ \citenamefont {Wineland}}]{Home2011Jul}%
  \BibitemOpen
  \bibfield  {author} {\bibinfo {author} {\bibfnamefont {J.~P.}\ \bibnamefont
  {Home}}, \bibinfo {author} {\bibfnamefont {D.}~\bibnamefont {Hanneke}},
  \bibinfo {author} {\bibfnamefont {J.~D.}\ \bibnamefont {Jost}}, \bibinfo
  {author} {\bibfnamefont {D.}~\bibnamefont {Leibfried}}, \ and\ \bibinfo
  {author} {\bibfnamefont {D.~J.}\ \bibnamefont {Wineland}},\ }\href {\doibase
  10.1088/1367-2630/13/7/073026} {\bibfield  {journal} {\bibinfo  {journal}
  {New J. Phys.}\ }\textbf {\bibinfo {volume} {13}},\ \bibinfo {pages} {073026}
  (\bibinfo {year} {2011})}\BibitemShut {NoStop}%
\bibitem [{\citenamefont {Wu}\ \emph {et~al.}(2018)\citenamefont {Wu},
  \citenamefont {Wang},\ and\ \citenamefont {Duan}}]{Wu2018Jun}%
  \BibitemOpen
  \bibfield  {author} {\bibinfo {author} {\bibfnamefont {Y.}~\bibnamefont
  {Wu}}, \bibinfo {author} {\bibfnamefont {S.-T.}\ \bibnamefont {Wang}}, \ and\
  \bibinfo {author} {\bibfnamefont {L.-M.}\ \bibnamefont {Duan}},\ }\href
  {\doibase 10.1103/PhysRevA.97.062325} {\bibfield  {journal} {\bibinfo
  {journal} {Phys. Rev. A}\ }\textbf {\bibinfo {volume} {97}},\ \bibinfo
  {pages} {062325} (\bibinfo {year} {2018})}\BibitemShut {NoStop}%
\bibitem [{\citenamefont {Alheit}\ \emph {et~al.}(1997)\citenamefont {Alheit},
  \citenamefont {Chu}, \citenamefont {Hoefer}, \citenamefont {Holzki},
  \citenamefont {Werth},\ and\ \citenamefont {Bl\"umel}}]{Alheit_1997}%
  \BibitemOpen
  \bibfield  {author} {\bibinfo {author} {\bibfnamefont {R.}~\bibnamefont
  {Alheit}}, \bibinfo {author} {\bibfnamefont {X.~Z.}\ \bibnamefont {Chu}},
  \bibinfo {author} {\bibfnamefont {M.}~\bibnamefont {Hoefer}}, \bibinfo
  {author} {\bibfnamefont {M.}~\bibnamefont {Holzki}}, \bibinfo {author}
  {\bibfnamefont {G.}~\bibnamefont {Werth}}, \ and\ \bibinfo {author}
  {\bibfnamefont {R.}~\bibnamefont {Bl\"umel}},\ }\href {\doibase
  10.1103/PhysRevA.56.4023} {\bibfield  {journal} {\bibinfo  {journal} {Phys.
  Rev. A}\ }\textbf {\bibinfo {volume} {56}},\ \bibinfo {pages} {4023}
  (\bibinfo {year} {1997})}\BibitemShut {NoStop}%
\bibitem [{\citenamefont {Razvi}\ \emph {et~al.}(1998)\citenamefont {Razvi},
  \citenamefont {Chu}, \citenamefont {Alheit}, \citenamefont {Werth},\ and\
  \citenamefont {Bl\"umel}}]{Razvi_1998}%
  \BibitemOpen
  \bibfield  {author} {\bibinfo {author} {\bibfnamefont {M.~A.~N.}\
  \bibnamefont {Razvi}}, \bibinfo {author} {\bibfnamefont {X.~Z.}\ \bibnamefont
  {Chu}}, \bibinfo {author} {\bibfnamefont {R.}~\bibnamefont {Alheit}},
  \bibinfo {author} {\bibfnamefont {G.}~\bibnamefont {Werth}}, \ and\ \bibinfo
  {author} {\bibfnamefont {R.}~\bibnamefont {Bl\"umel}},\ }\href {\doibase
  10.1103/PhysRevA.58.R34} {\bibfield  {journal} {\bibinfo  {journal} {Phys.
  Rev. A}\ }\textbf {\bibinfo {volume} {58}},\ \bibinfo {pages} {R34} (\bibinfo
  {year} {1998})}\BibitemShut {NoStop}%
\bibitem [{\citenamefont {Drakoudis}\ \emph {et~al.}(2006)\citenamefont
  {Drakoudis}, \citenamefont {Söllner},\ and\ \citenamefont
  {Werth}}]{Drakoudis_2006}%
  \BibitemOpen
  \bibfield  {author} {\bibinfo {author} {\bibfnamefont {A.}~\bibnamefont
  {Drakoudis}}, \bibinfo {author} {\bibfnamefont {M.}~\bibnamefont {Söllner}},
  \ and\ \bibinfo {author} {\bibfnamefont {G.}~\bibnamefont {Werth}},\ }\href
  {\doibase https://doi.org/10.1016/j.ijms.2006.02.006} {\bibfield  {journal}
  {\bibinfo  {journal} {International Journal of Mass Spectrometry}\ }\textbf
  {\bibinfo {volume} {252}},\ \bibinfo {pages} {61} (\bibinfo {year}
  {2006})}\BibitemShut {NoStop}%
\bibitem [{\citenamefont {Littich}(2011)}]{Littich}%
  \BibitemOpen
  \bibfield  {author} {\bibinfo {author} {\bibfnamefont {G.}~\bibnamefont
  {Littich}},\ }\emph {\bibinfo {title} {Electrostatic Control and Transport of
  Ions on a Planar Trap for. Quantum Information Processing}},\ \href@noop {}
  {Master's thesis},\ \bibinfo  {school} {ETH Zurich} (\bibinfo {year}
  {2011})\BibitemShut {NoStop}%
\bibitem [{\citenamefont {Sage}\ \emph {et~al.}(2012)\citenamefont {Sage},
  \citenamefont {Kerman},\ and\ \citenamefont {Chiaverini}}]{Sage2012Jul}%
  \BibitemOpen
  \bibfield  {author} {\bibinfo {author} {\bibfnamefont {J.~M.}\ \bibnamefont
  {Sage}}, \bibinfo {author} {\bibfnamefont {A.~J.}\ \bibnamefont {Kerman}}, \
  and\ \bibinfo {author} {\bibfnamefont {J.}~\bibnamefont {Chiaverini}},\
  }\href {\doibase 10.1103/PhysRevA.86.013417} {\bibfield  {journal} {\bibinfo
  {journal} {Phys. Rev. A}\ }\textbf {\bibinfo {volume} {86}},\ \bibinfo
  {pages} {013417} (\bibinfo {year} {2012})}\BibitemShut {NoStop}%
\bibitem [{\citenamefont {Dehmelt}(1968)}]{Dehmelt_1968}%
  \BibitemOpen
  \bibfield  {author} {\bibinfo {author} {\bibfnamefont {H.}~\bibnamefont
  {Dehmelt}},\ }\href@noop {} {\emph {\bibinfo {title} {Radiofrequency
  Spectroscopy of Stored Ions I: Storage}}},\ \bibinfo {series} {Advances in
  Atomic and Molecular Physics}, Vol.~\bibinfo {volume} {3}\ (\bibinfo
  {publisher} {Academic Press},\ \bibinfo {year} {1968})\ pp.\ \bibinfo {pages}
  {53--72}\BibitemShut {NoStop}%
\bibitem [{\citenamefont {Leibfried}\ \emph {et~al.}(2003)\citenamefont
  {Leibfried}, \citenamefont {Blatt}, \citenamefont {Monroe},\ and\
  \citenamefont {Wineland}}]{Leibfried_2003}%
  \BibitemOpen
  \bibfield  {author} {\bibinfo {author} {\bibfnamefont {D.}~\bibnamefont
  {Leibfried}}, \bibinfo {author} {\bibfnamefont {R.}~\bibnamefont {Blatt}},
  \bibinfo {author} {\bibfnamefont {C.}~\bibnamefont {Monroe}}, \ and\ \bibinfo
  {author} {\bibfnamefont {D.}~\bibnamefont {Wineland}},\ }\href@noop {}
  {\bibfield  {journal} {\bibinfo  {journal} {Rev. Mod. Phys.}\ }\textbf
  {\bibinfo {volume} {75}},\ \bibinfo {pages} {281} (\bibinfo {year}
  {2003})}\BibitemShut {NoStop}%
\bibitem [{\citenamefont {Xu}\ \emph {et~al.}(2023)\citenamefont {Xu},
  \citenamefont {Xia}, \citenamefont {Yu}, \citenamefont {Khan}, \citenamefont
  {Megidish}, \citenamefont {You}, \citenamefont {Hemmerling}, \citenamefont
  {Jayich}, \citenamefont {Biener},\ and\ \citenamefont {Häffner}}]{Xu_2023}%
  \BibitemOpen
  \bibfield  {author} {\bibinfo {author} {\bibfnamefont {S.}~\bibnamefont
  {Xu}}, \bibinfo {author} {\bibfnamefont {X.}~\bibnamefont {Xia}}, \bibinfo
  {author} {\bibfnamefont {Q.}~\bibnamefont {Yu}}, \bibinfo {author}
  {\bibfnamefont {S.}~\bibnamefont {Khan}}, \bibinfo {author} {\bibfnamefont
  {E.}~\bibnamefont {Megidish}}, \bibinfo {author} {\bibfnamefont
  {B.}~\bibnamefont {You}}, \bibinfo {author} {\bibfnamefont {B.}~\bibnamefont
  {Hemmerling}}, \bibinfo {author} {\bibfnamefont {A.}~\bibnamefont {Jayich}},
  \bibinfo {author} {\bibfnamefont {J.}~\bibnamefont {Biener}}, \ and\ \bibinfo
  {author} {\bibfnamefont {H.}~\bibnamefont {Häffner}},\ }\href@noop {}
  {\bibfield  {journal} {\bibinfo  {journal} {arXiv}\ } (\bibinfo {year}
  {2023})},\ \Eprint {http://arxiv.org/abs/2310.00595} {2310.00595 [quant-ph]}
  \BibitemShut {NoStop}%
\bibitem [{\citenamefont {Sterling}\ \emph {et~al.}(2013)\citenamefont
  {Sterling}, \citenamefont {Hughes}, \citenamefont {Mellor},\ and\
  \citenamefont {Hensinger}}]{Sterling_2013}%
  \BibitemOpen
  \bibfield  {author} {\bibinfo {author} {\bibfnamefont {R.~C.}\ \bibnamefont
  {Sterling}}, \bibinfo {author} {\bibfnamefont {M.~D.}\ \bibnamefont
  {Hughes}}, \bibinfo {author} {\bibfnamefont {C.~J.}\ \bibnamefont {Mellor}},
  \ and\ \bibinfo {author} {\bibfnamefont {W.~K.}\ \bibnamefont {Hensinger}},\
  }\href@noop {} {\bibfield  {journal} {\bibinfo  {journal} {Applied Physics
  Letters}\ }\textbf {\bibinfo {volume} {103}} (\bibinfo {year}
  {2013})}\BibitemShut {NoStop}%
\bibitem [{\citenamefont {Ray}\ \emph {et~al.}(2019)\citenamefont {Ray},
  \citenamefont {Rubenstein}, \citenamefont {Gu},\ and\ \citenamefont
  {Lordi}}]{Ray_2019}%
  \BibitemOpen
  \bibfield  {author} {\bibinfo {author} {\bibfnamefont {K.~G.}\ \bibnamefont
  {Ray}}, \bibinfo {author} {\bibfnamefont {B.~M.}\ \bibnamefont {Rubenstein}},
  \bibinfo {author} {\bibfnamefont {W.}~\bibnamefont {Gu}}, \ and\ \bibinfo
  {author} {\bibfnamefont {V.}~\bibnamefont {Lordi}},\ }\href {\doibase
  10.1088/1367-2630/ab1875} {\bibfield  {journal} {\bibinfo  {journal} {New
  Journal of Physics}\ }\textbf {\bibinfo {volume} {21}},\ \bibinfo {pages}
  {053043} (\bibinfo {year} {2019})}\BibitemShut {NoStop}%
\bibitem [{\citenamefont {Hite}\ \emph {et~al.}(2021)\citenamefont {Hite},
  \citenamefont {McKay},\ and\ \citenamefont {Pappas}}]{Hite_2021}%
  \BibitemOpen
  \bibfield  {author} {\bibinfo {author} {\bibfnamefont {D.~A.}\ \bibnamefont
  {Hite}}, \bibinfo {author} {\bibfnamefont {K.~S.}\ \bibnamefont {McKay}}, \
  and\ \bibinfo {author} {\bibfnamefont {D.~P.}\ \bibnamefont {Pappas}},\
  }\href {\doibase 10.1088/1367-2630/ac2c2c} {\bibfield  {journal} {\bibinfo
  {journal} {New Journal of Physics}\ }\textbf {\bibinfo {volume} {23}}
  (\bibinfo {year} {2021}),\ 10.1088/1367-2630/ac2c2c}\BibitemShut {NoStop}%
\bibitem [{\citenamefont {Cirac}\ and\ \citenamefont
  {Zoller}(2000)}]{Cirac_2000}%
  \BibitemOpen
  \bibfield  {author} {\bibinfo {author} {\bibfnamefont {J.~I.}\ \bibnamefont
  {Cirac}}\ and\ \bibinfo {author} {\bibfnamefont {P.}~\bibnamefont {Zoller}},\
  }\href@noop {} {\bibfield  {journal} {\bibinfo  {journal} {Nature}\ }
  (\bibinfo {year} {2000})}\BibitemShut {NoStop}%
\bibitem [{\citenamefont {Meinelt}\ \emph {et~al.}(2024)\citenamefont
  {Meinelt}, \citenamefont {Bahr}, \citenamefont {Finnegan}, \citenamefont
  {Haltli}, \citenamefont {Jordan}, \citenamefont {Klitsner}, \citenamefont
  {Liebsch}, \citenamefont {Mounce}, \citenamefont {Weatherred},\ and\
  \citenamefont {Stick}}]{Meinelt_2024}%
  \BibitemOpen
  \bibfield  {author} {\bibinfo {author} {\bibfnamefont {Z.~K.}\ \bibnamefont
  {Meinelt}}, \bibinfo {author} {\bibfnamefont {M.}~\bibnamefont {Bahr}},
  \bibinfo {author} {\bibfnamefont {P.~S.}\ \bibnamefont {Finnegan}}, \bibinfo
  {author} {\bibfnamefont {R.~A.}\ \bibnamefont {Haltli}}, \bibinfo {author}
  {\bibfnamefont {M.}~\bibnamefont {Jordan}}, \bibinfo {author} {\bibfnamefont
  {B.~H.}\ \bibnamefont {Klitsner}}, \bibinfo {author} {\bibfnamefont
  {T.}~\bibnamefont {Liebsch}}, \bibinfo {author} {\bibfnamefont
  {A.}~\bibnamefont {Mounce}}, \bibinfo {author} {\bibfnamefont {S.~E.}\
  \bibnamefont {Weatherred}}, \ and\ \bibinfo {author} {\bibfnamefont {D.~L.}\
  \bibnamefont {Stick}},\ }\href {\doibase 10.2172/2462919} {\emph {\bibinfo
  {title} {Microfabricated Ion Traps on Sapphire for Larger Trap Areas and
  Higher Qubit Count}}},\ \bibinfo {type} {Tech. Rep.}\ (\bibinfo
  {institution} {Sandia National Lab. (SNL-NM), Albuquerque, NM (United
  States)},\ \bibinfo {year} {2024})\BibitemShut {NoStop}%
\bibitem [{\citenamefont {Amini}\ \emph {et~al.}(2010)\citenamefont {Amini},
  \citenamefont {Uys}, \citenamefont {Wesenberg}, \citenamefont {Seidelin},
  \citenamefont {Britton}, \citenamefont {Bollinger}, \citenamefont
  {Leibfried}, \citenamefont {Ospelkaus}, \citenamefont {VanDevender},\ and\
  \citenamefont {Wineland}}]{Amini2010Mar}%
  \BibitemOpen
  \bibfield  {author} {\bibinfo {author} {\bibfnamefont {J.~M.}\ \bibnamefont
  {Amini}}, \bibinfo {author} {\bibfnamefont {H.}~\bibnamefont {Uys}}, \bibinfo
  {author} {\bibfnamefont {J.~H.}\ \bibnamefont {Wesenberg}}, \bibinfo {author}
  {\bibfnamefont {S.}~\bibnamefont {Seidelin}}, \bibinfo {author}
  {\bibfnamefont {J.}~\bibnamefont {Britton}}, \bibinfo {author} {\bibfnamefont
  {J.~J.}\ \bibnamefont {Bollinger}}, \bibinfo {author} {\bibfnamefont
  {D.}~\bibnamefont {Leibfried}}, \bibinfo {author} {\bibfnamefont
  {C.}~\bibnamefont {Ospelkaus}}, \bibinfo {author} {\bibfnamefont {A.~P.}\
  \bibnamefont {VanDevender}}, \ and\ \bibinfo {author} {\bibfnamefont {D.~J.}\
  \bibnamefont {Wineland}},\ }\href {\doibase 10.1088/1367-2630/12/3/033031}
  {\bibfield  {journal} {\bibinfo  {journal} {New J. Phys.}\ }\textbf {\bibinfo
  {volume} {12}},\ \bibinfo {pages} {033031} (\bibinfo {year}
  {2010})}\BibitemShut {NoStop}%
\bibitem [{\citenamefont {Burton}\ \emph {et~al.}(2023)\citenamefont {Burton},
  \citenamefont {Estey}, \citenamefont {Hoffman}, \citenamefont {Perry},
  \citenamefont {Volin},\ and\ \citenamefont {Price}}]{Burton2023Apr}%
  \BibitemOpen
  \bibfield  {author} {\bibinfo {author} {\bibfnamefont {W.~C.}\ \bibnamefont
  {Burton}}, \bibinfo {author} {\bibfnamefont {B.}~\bibnamefont {Estey}},
  \bibinfo {author} {\bibfnamefont {I.~M.}\ \bibnamefont {Hoffman}}, \bibinfo
  {author} {\bibfnamefont {A.~R.}\ \bibnamefont {Perry}}, \bibinfo {author}
  {\bibfnamefont {C.}~\bibnamefont {Volin}}, \ and\ \bibinfo {author}
  {\bibfnamefont {G.}~\bibnamefont {Price}},\ }\href {\doibase
  10.1103/PhysRevLett.130.173202} {\bibfield  {journal} {\bibinfo  {journal}
  {Phys. Rev. Lett.}\ }\textbf {\bibinfo {volume} {130}},\ \bibinfo {pages}
  {173202} (\bibinfo {year} {2023})}\BibitemShut {NoStop}%
\bibitem [{\citenamefont {Shu}\ \emph {et~al.}(2014)\citenamefont {Shu},
  \citenamefont {Vittorini}, \citenamefont {Buikema}, \citenamefont {Nichols},
  \citenamefont {Volin}, \citenamefont {Stick},\ and\ \citenamefont
  {Brown}}]{Shu2014Jun}%
  \BibitemOpen
  \bibfield  {author} {\bibinfo {author} {\bibfnamefont {G.}~\bibnamefont
  {Shu}}, \bibinfo {author} {\bibfnamefont {G.}~\bibnamefont {Vittorini}},
  \bibinfo {author} {\bibfnamefont {A.}~\bibnamefont {Buikema}}, \bibinfo
  {author} {\bibfnamefont {C.~S.}\ \bibnamefont {Nichols}}, \bibinfo {author}
  {\bibfnamefont {C.}~\bibnamefont {Volin}}, \bibinfo {author} {\bibfnamefont
  {D.}~\bibnamefont {Stick}}, \ and\ \bibinfo {author} {\bibfnamefont {K.~R.}\
  \bibnamefont {Brown}},\ }\href {\doibase 10.1103/PhysRevA.89.062308}
  {\bibfield  {journal} {\bibinfo  {journal} {Phys. Rev. A}\ }\textbf {\bibinfo
  {volume} {89}},\ \bibinfo {pages} {062308} (\bibinfo {year}
  {2014})}\BibitemShut {NoStop}%
\bibitem [{\citenamefont {Decaroli}\ \emph {et~al.}(2021)\citenamefont
  {Decaroli} \emph {et~al.}}]{Decaroli_2021}%
  \BibitemOpen
  \bibfield  {author} {\bibinfo {author} {\bibfnamefont {A.}~\bibnamefont
  {Decaroli}} \emph {et~al.},\ }\href@noop {} {\bibfield  {journal} {\bibinfo
  {journal} {Quantum Science and Technology}\ }\textbf {\bibinfo {volume}
  {6}},\ \bibinfo {pages} {044001} (\bibinfo {year} {2021})}\BibitemShut
  {NoStop}%
\bibitem [{\citenamefont {Ragg}\ \emph {et~al.}(2019)\citenamefont {Ragg},
  \citenamefont {Decaroli}, \citenamefont {Lutz},\ and\ \citenamefont
  {Home}}]{Ragg_2019}%
  \BibitemOpen
  \bibfield  {author} {\bibinfo {author} {\bibfnamefont {S.}~\bibnamefont
  {Ragg}}, \bibinfo {author} {\bibfnamefont {C.}~\bibnamefont {Decaroli}},
  \bibinfo {author} {\bibfnamefont {T.}~\bibnamefont {Lutz}}, \ and\ \bibinfo
  {author} {\bibfnamefont {J.~P.}\ \bibnamefont {Home}},\ }\href {\doibase
  10.1063/1.5119785} {\bibfield  {journal} {\bibinfo  {journal} {Review of
  Scientific Instruments}\ }\textbf {\bibinfo {volume} {90}} (\bibinfo {year}
  {2019}),\ 10.1063/1.5119785}\BibitemShut {NoStop}%
\bibitem [{\citenamefont {Wilpers}\ \emph {et~al.}(2012)\citenamefont
  {Wilpers}, \citenamefont {See}, \citenamefont {Gill},\ and\ \citenamefont
  {Sinclair}}]{Wilpers_2012}%
  \BibitemOpen
  \bibfield  {author} {\bibinfo {author} {\bibfnamefont {G.}~\bibnamefont
  {Wilpers}}, \bibinfo {author} {\bibfnamefont {P.}~\bibnamefont {See}},
  \bibinfo {author} {\bibfnamefont {P.}~\bibnamefont {Gill}}, \ and\ \bibinfo
  {author} {\bibfnamefont {A.~G.}\ \bibnamefont {Sinclair}},\ }\href {\doibase
  10.1038/NNANO.2012.126} {\bibfield  {journal} {\bibinfo  {journal} {Nature
  Nanotechnology}\ }\textbf {\bibinfo {volume} {7}},\ \bibinfo {pages} {572}
  (\bibinfo {year} {2012})}\BibitemShut {NoStop}%
\bibitem [{\citenamefont {Park}\ \emph {et~al.}(2024)\citenamefont {Park},
  \citenamefont {Notaros}, \citenamefont {Mohanty}, \citenamefont {Kim},
  \citenamefont {Notaros},\ and\ \citenamefont {Mouradian}}]{Park2024Sep}%
  \BibitemOpen
  \bibfield  {author} {\bibinfo {author} {\bibfnamefont {S.}~\bibnamefont
  {Park}}, \bibinfo {author} {\bibfnamefont {M.}~\bibnamefont {Notaros}},
  \bibinfo {author} {\bibfnamefont {A.}~\bibnamefont {Mohanty}}, \bibinfo
  {author} {\bibfnamefont {D.}~\bibnamefont {Kim}}, \bibinfo {author}
  {\bibfnamefont {J.}~\bibnamefont {Notaros}}, \ and\ \bibinfo {author}
  {\bibfnamefont {S.}~\bibnamefont {Mouradian}},\ }\href {\doibase
  10.1016/j.pquantelec.2024.100534} {\bibfield  {journal} {\bibinfo  {journal}
  {Prog. Quantum Electron.}\ }\textbf {\bibinfo {volume} {97}},\ \bibinfo
  {pages} {100534} (\bibinfo {year} {2024})}\BibitemShut {NoStop}%
\bibitem [{\citenamefont {Delaney}\ \emph {et~al.}(2024)\citenamefont
  {Delaney}, \citenamefont {Sletten}, \citenamefont {Cich}, \citenamefont
  {Estey}, \citenamefont {Fabrikant}, \citenamefont {Hayes}, \citenamefont
  {Hoffman}, \citenamefont {Hostetter}, \citenamefont {Langer}, \citenamefont
  {Moses}, \citenamefont {Perry}, \citenamefont {Peterson}, \citenamefont
  {Schaffer}, \citenamefont {Volin}, \citenamefont {Vittorini},\ and\
  \citenamefont {Burton}}]{Delaney_2024}%
  \BibitemOpen
  \bibfield  {author} {\bibinfo {author} {\bibfnamefont {R.~D.}\ \bibnamefont
  {Delaney}}, \bibinfo {author} {\bibfnamefont {L.~R.}\ \bibnamefont
  {Sletten}}, \bibinfo {author} {\bibfnamefont {M.~J.}\ \bibnamefont {Cich}},
  \bibinfo {author} {\bibfnamefont {B.}~\bibnamefont {Estey}}, \bibinfo
  {author} {\bibfnamefont {M.~I.}\ \bibnamefont {Fabrikant}}, \bibinfo {author}
  {\bibfnamefont {D.}~\bibnamefont {Hayes}}, \bibinfo {author} {\bibfnamefont
  {I.~M.}\ \bibnamefont {Hoffman}}, \bibinfo {author} {\bibfnamefont
  {J.}~\bibnamefont {Hostetter}}, \bibinfo {author} {\bibfnamefont
  {C.}~\bibnamefont {Langer}}, \bibinfo {author} {\bibfnamefont {S.~A.}\
  \bibnamefont {Moses}}, \bibinfo {author} {\bibfnamefont {A.~R.}\ \bibnamefont
  {Perry}}, \bibinfo {author} {\bibfnamefont {T.~A.}\ \bibnamefont {Peterson}},
  \bibinfo {author} {\bibfnamefont {A.}~\bibnamefont {Schaffer}}, \bibinfo
  {author} {\bibfnamefont {C.}~\bibnamefont {Volin}}, \bibinfo {author}
  {\bibfnamefont {G.}~\bibnamefont {Vittorini}}, \ and\ \bibinfo {author}
  {\bibfnamefont {W.~C.}\ \bibnamefont {Burton}},\ }\href {\doibase
  10.1103/physrevx.14.041028} {\bibfield  {journal} {\bibinfo  {journal}
  {Physical Review X}\ }\textbf {\bibinfo {volume} {14}} (\bibinfo {year}
  {2024}),\ 10.1103/physrevx.14.041028}\BibitemShut {NoStop}%
\bibitem [{\citenamefont {Mehta}\ \emph {et~al.}(2020)\citenamefont {Mehta},
  \citenamefont {Zhang}, \citenamefont {Malinowski}, \citenamefont {Nguyen},
  \citenamefont {Stadler},\ and\ \citenamefont {Home}}]{Mehta2020Oct}%
  \BibitemOpen
  \bibfield  {author} {\bibinfo {author} {\bibfnamefont {K.~K.}\ \bibnamefont
  {Mehta}}, \bibinfo {author} {\bibfnamefont {C.}~\bibnamefont {Zhang}},
  \bibinfo {author} {\bibfnamefont {M.}~\bibnamefont {Malinowski}}, \bibinfo
  {author} {\bibfnamefont {T.-L.}\ \bibnamefont {Nguyen}}, \bibinfo {author}
  {\bibfnamefont {M.}~\bibnamefont {Stadler}}, \ and\ \bibinfo {author}
  {\bibfnamefont {J.~P.}\ \bibnamefont {Home}},\ }\href {\doibase
  10.1038/s41586-020-2823-6} {\bibfield  {journal} {\bibinfo  {journal}
  {Nature}\ }\textbf {\bibinfo {volume} {586}},\ \bibinfo {pages} {533}
  (\bibinfo {year} {2020})}\BibitemShut {NoStop}%
\bibitem [{\citenamefont {Niffenegger}\ \emph {et~al.}(2020)\citenamefont
  {Niffenegger}, \citenamefont {Stuart}, \citenamefont {Sorace-Agaskar},
  \citenamefont {Kharas}, \citenamefont {Bramhavar}, \citenamefont {Bruzewicz},
  \citenamefont {Loh}, \citenamefont {Maxson}, \citenamefont {McConnell},
  \citenamefont {Reens}, \citenamefont {West}, \citenamefont {Sage},\ and\
  \citenamefont {Chiaverini}}]{Niffenegger2020Oct}%
  \BibitemOpen
  \bibfield  {author} {\bibinfo {author} {\bibfnamefont {R.~J.}\ \bibnamefont
  {Niffenegger}}, \bibinfo {author} {\bibfnamefont {J.}~\bibnamefont {Stuart}},
  \bibinfo {author} {\bibfnamefont {C.}~\bibnamefont {Sorace-Agaskar}},
  \bibinfo {author} {\bibfnamefont {D.}~\bibnamefont {Kharas}}, \bibinfo
  {author} {\bibfnamefont {S.}~\bibnamefont {Bramhavar}}, \bibinfo {author}
  {\bibfnamefont {C.~D.}\ \bibnamefont {Bruzewicz}}, \bibinfo {author}
  {\bibfnamefont {W.}~\bibnamefont {Loh}}, \bibinfo {author} {\bibfnamefont
  {R.~T.}\ \bibnamefont {Maxson}}, \bibinfo {author} {\bibfnamefont
  {R.}~\bibnamefont {McConnell}}, \bibinfo {author} {\bibfnamefont
  {D.}~\bibnamefont {Reens}}, \bibinfo {author} {\bibfnamefont {G.~N.}\
  \bibnamefont {West}}, \bibinfo {author} {\bibfnamefont {J.~M.}\ \bibnamefont
  {Sage}}, \ and\ \bibinfo {author} {\bibfnamefont {J.}~\bibnamefont
  {Chiaverini}},\ }\href {\doibase 10.1038/s41586-020-2811-x} {\bibfield
  {journal} {\bibinfo  {journal} {Nature}\ }\textbf {\bibinfo {volume} {586}},\
  \bibinfo {pages} {538} (\bibinfo {year} {2020})}\BibitemShut {NoStop}%
\bibitem [{\citenamefont {Shirao}\ \emph {et~al.}(2022)\citenamefont {Shirao},
  \citenamefont {Klawson}, \citenamefont {Mouradian},\ and\ \citenamefont
  {Wu}}]{Shirao2022Jun}%
  \BibitemOpen
  \bibfield  {author} {\bibinfo {author} {\bibfnamefont {M.}~\bibnamefont
  {Shirao}}, \bibinfo {author} {\bibfnamefont {D.}~\bibnamefont {Klawson}},
  \bibinfo {author} {\bibfnamefont {S.}~\bibnamefont {Mouradian}}, \ and\
  \bibinfo {author} {\bibfnamefont {M.~C.}\ \bibnamefont {Wu}},\ }\href
  {\doibase 10.35848/1347-4065/ac5b27} {\bibfield  {journal} {\bibinfo
  {journal} {Jpn. J. Appl. Phys.}\ }\textbf {\bibinfo {volume} {61}},\ \bibinfo
  {pages} {SK1002} (\bibinfo {year} {2022})}\BibitemShut {NoStop}%
\bibitem [{\citenamefont {Hensinger}\ \emph {et~al.}(2006)\citenamefont
  {Hensinger}, \citenamefont {Olmschenk}, \citenamefont {Stick}, \citenamefont
  {Hucul}, \citenamefont {Yeo}, \citenamefont {Acton}, \citenamefont
  {Deslauriers}, \citenamefont {Monroe},\ and\ \citenamefont
  {Rabchuk}}]{Hensinger2006Jan}%
  \BibitemOpen
  \bibfield  {author} {\bibinfo {author} {\bibfnamefont {W.~K.}\ \bibnamefont
  {Hensinger}}, \bibinfo {author} {\bibfnamefont {S.}~\bibnamefont
  {Olmschenk}}, \bibinfo {author} {\bibfnamefont {D.}~\bibnamefont {Stick}},
  \bibinfo {author} {\bibfnamefont {D.}~\bibnamefont {Hucul}}, \bibinfo
  {author} {\bibfnamefont {M.}~\bibnamefont {Yeo}}, \bibinfo {author}
  {\bibfnamefont {M.}~\bibnamefont {Acton}}, \bibinfo {author} {\bibfnamefont
  {L.}~\bibnamefont {Deslauriers}}, \bibinfo {author} {\bibfnamefont
  {C.}~\bibnamefont {Monroe}}, \ and\ \bibinfo {author} {\bibfnamefont
  {J.}~\bibnamefont {Rabchuk}},\ }\href {\doibase 10.1063/1.2164910} {\bibfield
   {journal} {\bibinfo  {journal} {Appl. Phys. Lett.}\ }\textbf {\bibinfo
  {volume} {88}},\ \bibinfo {pages} {034101} (\bibinfo {year}
  {2006})}\BibitemShut {NoStop}%
\bibitem [{\citenamefont {Blakestad}\ \emph {et~al.}(2009)\citenamefont
  {Blakestad}, \citenamefont {Ospelkaus}, \citenamefont {VanDevender},
  \citenamefont {Amini}, \citenamefont {Britton}, \citenamefont {Leibfried},\
  and\ \citenamefont {Wineland}}]{Blakestad2009Apr}%
  \BibitemOpen
  \bibfield  {author} {\bibinfo {author} {\bibfnamefont {R.~B.}\ \bibnamefont
  {Blakestad}}, \bibinfo {author} {\bibfnamefont {C.}~\bibnamefont
  {Ospelkaus}}, \bibinfo {author} {\bibfnamefont {A.~P.}\ \bibnamefont
  {VanDevender}}, \bibinfo {author} {\bibfnamefont {J.~M.}\ \bibnamefont
  {Amini}}, \bibinfo {author} {\bibfnamefont {J.}~\bibnamefont {Britton}},
  \bibinfo {author} {\bibfnamefont {D.}~\bibnamefont {Leibfried}}, \ and\
  \bibinfo {author} {\bibfnamefont {D.~J.}\ \bibnamefont {Wineland}},\ }\href
  {\doibase 10.1103/PhysRevLett.102.153002} {\bibfield  {journal} {\bibinfo
  {journal} {Phys. Rev. Lett.}\ }\textbf {\bibinfo {volume} {102}},\ \bibinfo
  {pages} {153002} (\bibinfo {year} {2009})}\BibitemShut {NoStop}%
\bibitem [{\citenamefont {Blakestad}\ \emph {et~al.}(2011)\citenamefont
  {Blakestad}, \citenamefont {Ospelkaus}, \citenamefont {VanDevender},
  \citenamefont {Wesenberg}, \citenamefont {Biercuk}, \citenamefont
  {Leibfried},\ and\ \citenamefont {Wineland}}]{Blakestad2011Sep}%
  \BibitemOpen
  \bibfield  {author} {\bibinfo {author} {\bibfnamefont {R.~B.}\ \bibnamefont
  {Blakestad}}, \bibinfo {author} {\bibfnamefont {C.}~\bibnamefont
  {Ospelkaus}}, \bibinfo {author} {\bibfnamefont {A.~P.}\ \bibnamefont
  {VanDevender}}, \bibinfo {author} {\bibfnamefont {J.~H.}\ \bibnamefont
  {Wesenberg}}, \bibinfo {author} {\bibfnamefont {M.~J.}\ \bibnamefont
  {Biercuk}}, \bibinfo {author} {\bibfnamefont {D.}~\bibnamefont {Leibfried}},
  \ and\ \bibinfo {author} {\bibfnamefont {D.~J.}\ \bibnamefont {Wineland}},\
  }\href {\doibase 10.1103/PhysRevA.84.032314} {\bibfield  {journal} {\bibinfo
  {journal} {Phys. Rev. A}\ }\textbf {\bibinfo {volume} {84}},\ \bibinfo
  {pages} {032314} (\bibinfo {year} {2011})}\BibitemShut {NoStop}%
\end{thebibliography}%

\end{document}